\documentclass[apj]{emulateapj}
\usepackage{amsmath, mathrsfs, amssymb}
\usepackage{apjfonts}
\usepackage{color}
\usepackage[colorlinks, citecolor=blue]{hyperref}

\def\gtrsim{\lower 2pt \hbox{$\, \buildrel {\scriptstyle >}\over
{\scriptstyle \sim}\,$}}
\def\lesssim{\lower 2pt \hbox{$\, \buildrel {\scriptstyle <}\over
{\scriptstyle \sim}\,$}}

\def\conunit{{\rm erg\,s^{-1}\,cm^{-2}\,\text{\AA}^{-1}}}
\def\lineunit{{\rm erg\,s^{-1}\,cm^{-2}}}

\def\bha{a_{\bullet}}
\def\bhe{e_{\bullet}}
\def\bhT{T_{\bullet}}
\def\bhm{M_{\bullet}}
\def\BHA{A_{\bullet}}

\def\kms{\rm km~s^{-1}}

\def\sunm{M_{\odot}}

\def\rhbeta{R_{\rm H\beta}}

\def\TFhb{T_{\rm F_{\rm H\beta}}}
\def\TPhb{T_{\rm P_{H\beta}}}

\def\oiii{{[\ion{O}{3}]}}
\def\feii{{\ion{Fe}{2}}}

\def\ihep{1}
\def\kiaa{2}
\def\pkudoa{3}
\def\bnu{5}
\def\ynao{6}
\def\naoc{4}

\slugcomment{To appear in {\it The Astrophysical Journal.}}
\lefthead{\sc{Li et al.}}
\righthead{\sc{Spectroscopic Indication for a Centi-parsec SMBHB in NGC~5548}}

\begin{document}

\title{Spectroscopic Indication for a Centi-parsec Supermassive Black 
Hole Binary in the Galactic Center of NGC~5548}

\author
{Yan-Rong Li,\altaffilmark{\ihep}
Jian-Min Wang,\altaffilmark{\ihep, \naoc, $\ast$}
Luis C. Ho,\altaffilmark{\kiaa, \pkudoa}
Kai-Xing Lu,\altaffilmark{\bnu, \ihep}
Jie Qiu,\altaffilmark{\ihep}
Pu Du,\altaffilmark{\ihep}
Chen Hu,\altaffilmark{\ihep}
Ying-Ke Huang,\altaffilmark{\ihep}
Zhi-Xiang Zhang,\altaffilmark{\ihep}
Kai Wang,\altaffilmark{\ihep}
and
Jin-Ming Bai $^{\ynao}$
}

\altaffiltext{1}{Key Laboratory for Particle Astrophysics, Institute of High Energy Physics, 
Chinese Academy of Sciences, Beijing 100049, China; wangjm@ihep.ac.cn}
\altaffiltext{2}{Kavli Institute of Astronomy and Astrophysics, Peking University, Beijing 
100871, China}
\altaffiltext{3}{Department of Astronomy, School of Physics, Peking University, Beijing 100871, China}
\altaffiltext{4}{National Astronomical Observatories of China, Chinese Academy of Sciences,
 Beijing 100020, China}
\altaffiltext{5}{Department of Astronomy, Beijing Normal University, Beijing 100875, China}
\altaffiltext{6}{Yunnan Observatories, Chinese Academy of Sciences, Kunming 650011, Yunnan, China}

\begin{abstract}
As a natural consequence of cosmological hierarchical structure formation,
sub-parsec supermassive black hole binaries (SMBHBs) should be common in 
galaxies but thus far have eluded spectroscopic identification.  
Based on four decades of optical spectroscopic 
monitoring, we report that the nucleus of NGC~5548, a nearby Seyfert 
galaxy long suspected to have experienced a major merger about one billion years 
ago, exhibits long-term variability with a period of $\sim$14 years in 
the optical continuum and broad H$\beta$ emission line. Remarkably, the double-peaked profile 
of H$\beta$ shows systematic velocity changes with a similar period.
These pieces of observations plausibly indicate that a SMBHB resides in
the center of NGC~5548.
The complex, secular variations in the line profiles can be explained by 
orbital motion of a binary with equal mass and a semi-major axis of $\sim22$ 
light-days (corresponding to $\sim$18 milli-parsec).  At a distance of 75~Mpc, 
NGC~5548 is one of the nearest sub-parsec SMBHB candidates that offers
an ideal laboratory for gravitational wave detection.

\end{abstract}

\section{Introduction}

There is increasing indirect evidence that supermassive black hole binaries 
(SMBHBs) are common in galactic centers, including the presence of cores in 
the central brightness distributions, which are regarded as a consequence of 
stars ejected by a SMBHB (\citealt{Ebisuzaki1991,Merritt2005,Kormendy2009}),
and long-term periodicity in optical light curves (\citealt{Valtonen2008,Graham2015,
Liu2015, Zheng2015}). Dual AGNs with kpc-scale separation, products of galaxy mergers, 
have been unambiguously detected in a number of cases (\citealt{Komossa2003,
Comerford2015,Fu2015}). However, the identification of sub-parsec SMBHBs is particularly 
challenging because these small separations at cosmic distance are well below 
the angular resolving power of the current and even most future powerful telescopes. 
Searching for the spectroscopic signatures is certainly an ideal solution 
(e.g., \citealt{Bon2012, Eracleous2012, Popovic2012, Shen2013, Liu2014, Yan2015} and 
references therein), although robust, conclusive evidence for sub-parsec SMBHBs 
remains rare. If both components of the binary contain a broad-line region (BLR), 
monitoring secular changes in the intensity and velocity of the line-emitting clouds 
offers a promising avenue to identify gravitationally bound, sub-parsec 
SMBHB pairs in their last phase of dynamical evolution.  

As one of the best-observed nearby AGNs, the Seyfert~1 galaxy 
NGC~5548 ($z=0.017175$) has been intensively monitored on short and long 
time scales, especially at optical/UV wavelengths (e.g., \citealt{Sergeev2007, 
Peterson2002, Iijima1995, Popovic2008, Bentz2009, Denney2010, DeRosa2015, Fausnaugh2015}).  
The earliest available data for the optical continuum date back to 1968 
(\citealt{Romanishin1995}), and for H$\beta$ back to 1972 (\citealt{Sergeev2007}), 
spanning a time range of over four decades. Its bright nucleus prevents us from 
getting a detailed view of the galaxy's central brightness distribution, 
but optical images show a classical bulge, a disk, and two prominent tidal 
features: a bright long tail bending around the galaxy (\citealt{Schweizer1988, MacKenty1990}) 
and a faint, long, straight tail only seen in very deep exposures 
(\citealt{Tyson1998}). In addition, the disk of NGC~5548 exhibits a distorted morphology with
partial ring structures or ripples (\citealt{Schweizer1988, Tyson1998}).
These imprints provide ``smoking-gun'' evidence of a major merger that occurred 
$\sim 1$ Gyr ago (see \citealt{Tyson1998} for a detailed discussion).  According to 
the standard scenario of SMBHB evolution (\citealt{Begelman1980,Merritt2005}), we expect 
the center of NGC~5548 to contain a tightly bound SMBHB pair.

\begin{figure*}[th!]
\centering
\includegraphics[angle=0,width=0.7\textwidth]{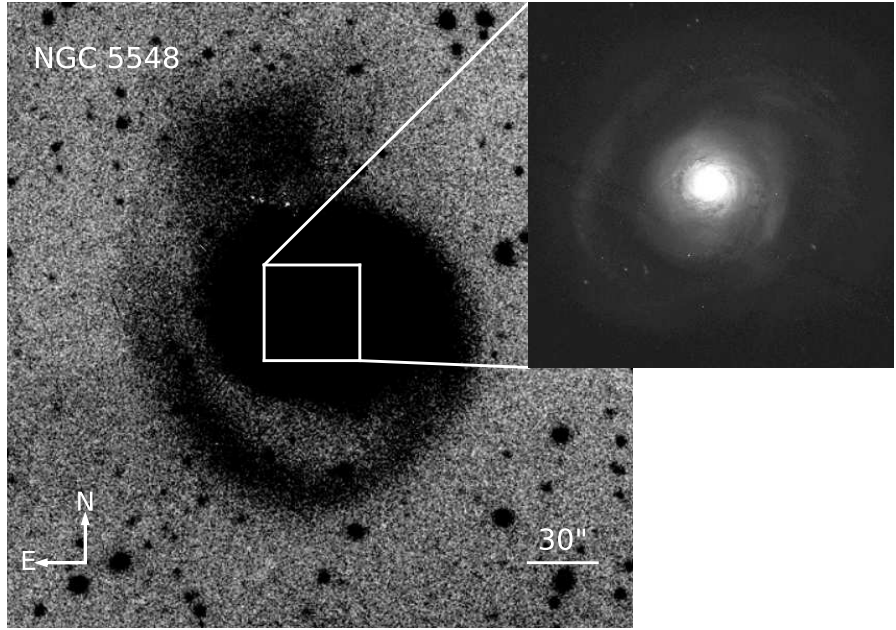}
\caption{\footnotesize
Optical image of NGC~5548. A deep, $V$-band image taken with the Lijiang 2.4m telescope 
showing a long, tidal tail wrapping around the galaxy, indicative of a major merger $\sim 1$ Gyr ago. 
The inset panel highlights small-scale structure visible in a WFC3/F547M image obtained from 
the {\it HST}\ archives.}
\label{fig_img}
\end{figure*}

Early in 1987, using data from their three-year spectroscopic monitoring, \cite{PK1987}
proposed that NGC~5548 contains a SMBHB to explain its asymmetric,  
varying H$\beta$ profiles. They roughly estimated that the binary period is 
of the order 110 yr.  Using four decades of optical spectroscopic monitoring, 
we provide compelling evidence for a SMBHB in the galactic center of NGC~5548 based on
more robust {\it measurements} of the period and binary orbital information. 

The paper is organized as follows. In Section 2, we summarize the historical 
data available in the literature and our new observations taken in 2015 using the Lijiang 
2.4m telescope and outline the procedures for homogeneous spectral measurements.
In Section 3, we exploit the periodicity in the long-term variations of the continuum
and H$\beta$ fluxes, and in  the H$\beta$ profile changes. We find that these three 
periods are remarkably all in agreement within their uncertainties. Most importantly, 
the double-peaked H$\beta$ profiles shift and merge in a periodic manner.
Together with the tidal tails of NGC~5548, these pieces of evidence indicate 
a SMBHB in its galactic center. In Section~4, we show that the lines of
observations presented here can test existing interpretations
for the periodicity in the continuum, H$\beta$ fluxes, and H$\beta$ profile 
changes, and the presence of a SMBHB is the most natural interpretation.
We introduce a simple toy model of BLRs around a SMBHB and fit the H$\beta$ 
profiles to extract the orbital information in Section~5. We discuss the 
possible caveats of our SMBHB interpretation and future improvements in Section~6. 
The conclusions are summarized in Section~7.

\begin{deluxetable*}{cccccl}
  \tablecolumns{5}
  \tablewidth{0.95\textwidth}
  \tablecaption{Spectroscopic datasets of NGC~5548.}
  \tabletypesize{\footnotesize}
\tablehead{
  \colhead{Dataset}                       &
  \colhead{Ref.}                          &
  \multicolumn{2}{c}{Observation Periods} &
  \colhead{Number of}                       &
  \colhead{Note}                            \\ \cline{3-4}
  \colhead{}                              &
  \colhead{}                              &
  \colhead{JD (-2,400,000)  }              &
  \colhead{Year }                         &
  \colhead{Spectra}                              &
  \colhead{}                             
 }
\startdata
S07     & 1          & 41,420$-$52,174  & 1972 - 2001 & 814     & Includes AGN Watch dataset\\
I95     & 2          & 47,169$-$49,534  & 1988$-$1994 & 23      & \nodata\\
P08     & 3          & 51,020$-$53,065  & 1998$-$2004 & 22      & \nodata\\
SDSS06  & \nodata    & 53,859           &  2006       &  1      & SDSS database  \\
AGN05   & 4          & 53,431$-$53,472  &  2005       & \nodata & Only H$\beta$ fluxes available \\   
AGN07   & 5          & 54,180$-$54,333  &  2007       &  1      & Mean spectrum \\
LAMP08  & 6          & 54,550$-$54,619  &  2008       &  1      & Mean spectrum \\
M15\tablenotemark{a} & 7  &     56, 490 &  2013       &  1      &  Only used for measuring H$\beta$ flux   \\
This work&\nodata    & 57,030$-$57,236  &  2015       &  62     & \nodata
\enddata
\tablerefs{ (1) \cite{Sergeev2007};   (2) \cite{Iijima1995};  (3) \cite{Popovic2008}; (4) \cite{Bentz2007};
            (5) \cite{Denney2010};    (6) \cite{Bentz2009};   (7) \cite{Mehdipour2015}.}
\tablenotetext{a}{We digitized Figure~5 of \cite{Mehdipour2015} to obtain the spectrum.}
\label{tab_hb}
\end{deluxetable*}

\begin{deluxetable*}{cccccccl}
 \tablecolumns{5}
 \tablewidth{0.95\textwidth}
 \tablecaption{Datasets for the 5100 {\AA} continuum fluxes of NGC~5548}
 \tabletypesize{\footnotesize}
\tablehead{
 \colhead{Dataset}                       &
 \colhead{Ref.}                          &
 \multicolumn{2}{c}{Observation Periods} &
 \colhead{Number of}                     &
 \colhead{Aperture}                     &
 \colhead{$F_{\rm gal}$\tablenotemark{a}}                 &                                          
 \colhead{Note}                          \\ \cline{3-4}
 \colhead{}                              &
 \colhead{}                              &
 \colhead{JD (-2,400,000)}               &
 \colhead{Year}                          & 
 \colhead{Observations}                  &
 \colhead{(arcsec)}                      &
 \colhead{}                              &
 \colhead{}
 }
\startdata
R95\tablenotemark{b}    &1       & 39,941$-$45,174  & 1968$-$1982 & 17   & \nodata          &\nodata         & $V$-band photometry\\
S07                     &2       & 41,420$-$52,174  & 1972$-$2001 & 814  & $5.0\times7.5$   & $3.75\pm0.38$  & Partially overlaps with AGN Watch dataset\\
LD93                    &3       & 43,216$-$48,856  & 1977$-$1992 &  82  & 7.15             & $7.08\pm0.19$  & $V$-band photometry\\
I95                     &4       & 47,169$-$49,534  & 1988$-$1994 & 23   & $3.0\times12.0$  & $2.77\pm0.18$  & \nodata\\
AGN Watch               &5       & 47,509$-$52,265  & 1989$-$2001 & 1548 & $5.0\times7.5$   & $3.75\pm0.38$  &\nodata\\
A14                     &6       & 50,883$-$55,734  & 1998$-$2013 &  10  & \nodata          & \nodata        &1350 \AA\ fluxes   \\
P08                     &7       & 51,020$-$53,065  & 1998$-$2004 & 22   & $2.5\times6.0$   & $1.73\pm0.17$  & \nodata\\
K14\tablenotemark{c}    &8       & 51,992$-$54,332  & 2001$-$2007 & 302  & 4.15             & $4.21\pm0.19$  & $V$-band fluxes\\
R12\tablenotemark{d}    &9       & 52,818$-$55,716  & 2003$-$2011 & 34   & \nodata          & \nodata        &$V$-band photometry\\
AGN05                   &10      & 53,431$-$53,472  &  2005       &  40  & $5.0\times12.75$ & $4.34\pm0.43$  &\nodata \\
SDSS06                  &\nodata & 53,859           &  2006       &  1   & 1.5              & $1.16\pm0.14$  & SDSS database  \\
AGN07                   &11      & 54,180$-$54,333  &  2007       &  220 & $5.0\times12.0$  & $4.27\pm0.43$  &\nodata \\
LAMP08                  &12      & 54,509$-$54,617  &  2008       &  57  & $4.0\times9.4$   & $3.54\pm0.35$  &\nodata \\
AGN12\tablenotemark{e}  &13      & 55,932$-$56,048  &  2012       &  90  & $5.0\times15.0$  &  $4.45\pm0.44$ & \nodata\\
E15                     &14      & 56,431$-$56,824  & 2013$-$2014 & 409  & 5                & $5.07\pm0.19$  &$V$-band fluxes \\
This work               &\nodata & 57,030$-$57,236  &  2015       &  62  & $2.5\times8.5$   & $2.04\pm0.17$  & \nodata
\enddata
\tablerefs{(1)  \cite{Romanishin1995}; (2)  \cite{Sergeev2007};  (3) \cite{Lyutyi1993};   (4)  \cite{Iijima1995};
           (5)  \cite{Peterson2002};   (6) \cite{Arav2014};      (7) \cite{Popovic2008};  (8)  \cite{Koshida2014};
           (9) \cite{Roberts2012};     (10)  \cite{Bentz2007};   (11) \cite{Denney2010};  (12)  \cite{Bentz2009};   
           (13)  \cite{Peterson2013};  (14) \cite{Edelson2015}.}

\tablenotetext{a}{$F_{\rm gal}$ is the host galaxy contribution to the continuum in units 
                  of $10^{-15}~{\rm erg~s^{-1}~cm^{-2}~\text{\AA}^{-1}}$.}
\tablenotetext{b}{The authors had subtracted the host galaxy contribution.}
\tablenotetext{c}{We shifted the obtained 5100\ {\AA} fluxes by $-1.30\times  10^{-15}~{\rm 
                  erg~s^{-1}~cm^{-2}~\text{\AA}^{-1}}$ to align the values with the other datasets.}
\tablenotetext{d}{The aperture size was not explicitly specified in the reference. We adopt a host galaxy flux 
                  $F_{\rm gal} = 4.67\times 10^{-15}~{\rm erg~s^{-1}~cm^{-2}~\text{\AA}^{-1}}$ to align the fluxes 
                   with the other datasets. }
\tablenotetext{e}{We digitized Figure~1 of \cite{Peterson2013} to obtain the fluxes.}
\label{tab_con}
\end{deluxetable*}

\section{Observations of NGC 5548 and data processing}

\subsection{New observations in 2015}
We made spectroscopic and photometric observations of NGC~5548 from 
January 7 to August 1, 2015, using the Lijiang 2.4m telescope, located in 
Yunnan Province, China, operated by Yunnan Observatories. The Yunnan 
Faint Object Spectrograph and Camera (YFOSC) is equipped with a 
back-illuminated 2048$\times$4608 pixel CCD covering a field-of-view of
$10^{\prime}\times10^{\prime}$. The photometric observations use a Johnson 
$V$ filter. Details of the telescope, YFOSC, observational setup, and data 
reduction can be found in \cite{Du2014}.  To achieve accurate absolute and 
relative flux calibration, we took advantage of the long-slit capability of 
YFOSC and simultaneously observed a nearby comparison star as a reference 
standard.  Absolute flux calibration of the comparison star is obtained 
using observations of spectrophotometric standards during nights of good 
weather conditions. Given the seeing of the observations 
($1.0^{\prime\prime}-2.5^{\prime\prime}$), the slit was fixed at a projected 
width of $2.5^{\prime\prime}$.  The grism provides a resolution of 92~\AA~${\rm mm^{-1}}$ 
(1.8 {\AA} per pixel) and covers the wavelength range $3800-7200$~\AA. The spectra 
were extracted using a uniform window of $8.5^{\prime\prime}$. In total, 
we obtained 61 photometric observations and 62 spectroscopic observations, with typical 
exposures of 5 and 20 min, respectively. The detailed dataset and reverberation analysis 
will be presented in a separate paper (\citealt{Lu2016}).

To obtain a deep image of NGC~5548, we aligned and stacked all the photometric data, 
resulting in a long-exposure image with a total integration time of 3950 s.  
Figure~\ref{fig_img} shows the final stacked image, in which the bending long tail,
first detected by \cite{Schweizer1988}, is clearly seen. We also superpose 
a WFC3/F547M image obtained from the {\it Hubble Space Telescope (HST)} archives to 
highlight small-scale structure of NGC~5548.

\subsection{Historical data and construction of a homogeneous database}
NGC~5548 has been observed many times individually (e.g., \citealt{Iijima1995, 
Shapovalova2004, Sergeev2007, Popovic2008}), and 16 monitoring campaigns have 
been devoted to reverberation mapping of the H$\beta$ emission 
line (e.g., \citealt{Peterson2002, Bentz2007, Bentz2009, Denney2010}). 
In Tables~\ref{tab_hb} and \ref{tab_con}, we compile spectroscopic and photometric
data available in the literature and list the corresponding references. 
Appendix A.1 gives the detailed information for the spectroscopic data sources. 
We also collected $V$-band photometric data and UV 1350 {\AA} data. 
Appendix A.2 shows how to convert continuum fluxes from these bands 
to 5100~{\AA}. Because the aperture sizes are different among the data sources,
we subtract the host galaxy contribution from the 5100~{\AA}
fluxes as described in the next section. 

The database constitutes a variety of sources with different 
aperture sizes, resolution, and wavelength and flux calibration; therefore, 
it is important to compile and calibrate them in a systematic fashion.
Fortunately, most spectroscopic data sources listed in Tables~\ref{tab_hb} and 
\ref{tab_con} were calibrated uniformly, in a manner similar to that employed 
for the  AGN Watch project, based on a 
constant [\ion{O}{3}] flux (see below). This allows us to directly use their 
published 5100~{\AA} continuum and integrated H$\beta$ fluxes. However, for 
H$\beta$ profiles, extra calibration is required because the reference 
[\ion{O}{3}] profiles among the data sources are generally different in spectral resolution and 
wavelength calibration. 

\begin{figure*}[th!]
\centering
\includegraphics[width=0.8\textwidth]{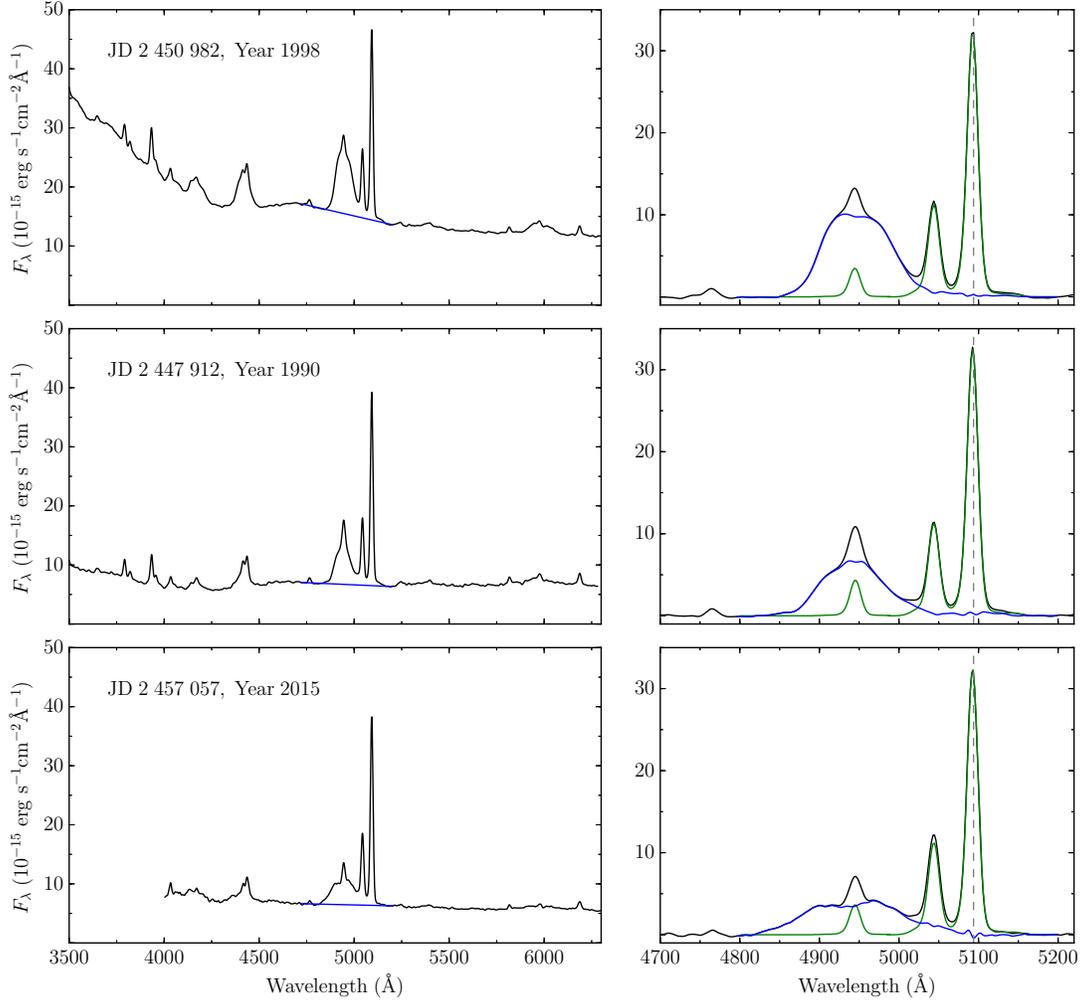}
\caption{Three illustrative examples for the spectral decomposition. Left panels: 
calibrated spectra, showing the blue straight line
used to subtract the pseudo continuum. Right panels: continuum-subtracted 
spectra around H$\beta$ line in the observed frame (see the text for details), 
where green line represents the narrow-line components (the narrow H$\beta$ and 
[\ion{O}{3}] doublet), blue line represents the broad H$\beta$ component, 
and grey dashed line represents the location of [\ion{O}{3}]~$\lambda5007$. Top 
two panels show the spectra at high and low state 
($F_{\rm 5100}=10.2, 3.1\times 10^{-15}~\conunit$), respectively, from the AGN 
Watch project, and bottom panel shows a spectrum from our new observations 
($F_{\rm 5100}=3.0\times 
10^{-15}~\conunit$).}
\label{fig_example}
\end{figure*}

The majority of the spectroscopic data comes from \cite{Sergeev2007} (S07; 
see Table~\ref{tab_hb}), who included archival data from 1972--1990 and high-quality 
data from the AGN Watch project from 1989--2001. The authors calibrated all their spectra 
based on the usual assumption that the narrow [\ion{O}{3}]~$\lambda5007$ line does not 
vary with time (\citealt{vanGroningen1992}) and further inter-calibrated the spectra 
to account for aperture effects (\citealt{Peterson1995, Li2014}). Inter-calibration 
is necessary because the narrow-line region of NGC~5548 is spatially resolved so that 
different apertures admit different amounts of light from the narrow-line region. 
The absolute flux of [\ion{O}{3}] was set to $F([\text{\ion{O}{3}}])=5.58\times10^{-13}\,\lineunit$
(\citealt{Peterson2002}). \citeauthor{Sergeev2007} measured the 5100~{\AA} continuum and decomposed the
broad H$\beta$ profiles with the narrow-line component subtracted, and kindly
made them available to us. We directly use these data for our analysis.  We align the spectra from
other data sources with those of S07, using the same data calibration 
processing. To be specific, by comparing the [\ion{O}{3}]~$\lambda5007$ line with 
respect to the dataset of S07, we correct for the differences of the other
datasets in spectral resolution, wavelength and flux calibration%
\footnote{We note that \cite{Peterson2013} showed that the [\ion{O}{3}] emission-line
flux of NGC~5548 has a long-term secular variation. We do not include such variation in 
our calibration because of two facts. First, there were no reliable absolute flux measurements 
of [\ion{O}{3}] line before 1988.  Second, the variation amplitude of of [\ion{O}{3}] 
over 1988--2012 is indeed small compared with that of the 5100~{\AA} continuum during the same period.
The flux change of of [\ion{O}{3}] was generally less than 10\%, and the largest 
change was $\sim 23\%$ in 2008, which can result in a shift in flux density at 5100~{\AA} of 
$\sim 0.5\times10^{-15}~\conunit$. The overall mean flux density of NGC~5548 at 5100~{\AA} is 
$\sim 10\times10^{-15}~\conunit$, indicating that the secular variation of [\ion{O}{3}] has an 
insignificant influence on our present analysis. Similarly, this secular variation of [\ion{O}{3}] 
is also unimportant to our analysis of the H$\beta$ fluxes.}.
We then perform inter-calibration for data sources I95 and P08 in Table~\ref{tab_hb} by
comparing the measurements of 5100~{\AA} continuum and H$\beta$ fluxes with those 
from S07 and AGN Watch that are closely spaced within a
time interval of 20 days. For data sources that do not overlap with the other sources
(e.g., those after 2004 in Table~\ref{tab_hb}), provided that the adopted aperture is
large enough (e.g., circular aperture radius $\gtrsim3^{\prime\prime}$), all the narrow-line 
flux should be enclosed so that inter-calibration is no longer needed; otherwise, we use the 
surface-brightness distribution of narrow [\ion{O}{3}]~$\lambda5007$ derived by \cite{Peterson1995}
to determine a correction factor. It turns out that only the data from source SDSS06 and from
this work require a correction.

Note that the AGN Watch project{\footnote{\url{http://www.astronomy.ohio-state.edu/~agnwatch}.}}
published their full data for 5100~{\AA} continuum and integrated H$\beta$ fluxes 
in \cite{Peterson2002}, and \cite{Sergeev2007} included only the high-quality spectra of 
the AGN Watch project in their compilation. Therefore, these two data sources partially 
overlap. For the overlapping portion, when analyzing the light curves of the 5100~{\AA} and 
integrated H$\beta$ fluxes, we use the full data from the AGN Watch project; whereas when 
analyzing the H$\beta$ profiles, which requires higher quality data, we only use the data from 
\cite{Sergeev2007}.

\subsection{Spectral measurements}
For the purpose of homogeneous spectral measurements,
we measure the 5100~{\AA} flux density 
and decompose the broad H$\beta$ line using the same standard procedure as in 
\cite{Sergeev2007}.  Specifically, we measure the 5100 {\AA} continuum flux $F_{5100}$ as the average 
flux density in a 50 \AA-wide wavelength range centered on 5100~\AA\ in the rest frame.  
The uncertainty is set by the standard deviation of fluxes in the same wavelength range. 
To correct for host galaxy contribution to the 5100~{\AA} continuum, we use the $V$-band 
aperture magnitudes of the host galaxy derived by \cite{Romanishin1995}. The 
$V$-band magnitude is converted into 5100~{\AA} continuum flux ($F_{5100}$)
by assuming that NGC~5548 has a galaxy spectral template similar to that of M32. 
This leads to $F_{5100}=3.4\times10^{-15}\,\conunit$ for a $V$-band magnitude of 
14.99 (\citealt{Romanishin1995}).  We compared our host galaxy corrections with those 
given in \cite{Peterson2013} and find that the differences are generally smaller than 10\%. 

To decompose the broad H$\beta$ line, we subtract the underlying continuum, the narrow 
H$\beta$ component, and the \oiii$~\lambda\lambda$4959, 5007 doublet, which severely blends 
with H$\beta$ during the low-flux states of NGC~5548. Since the spectra have been calibrated 
to a common [\ion{O}{3}]~$\lambda5007$ flux, the doublet can be subtracted uniformly by 
creating a template. We make use of the [\ion{O}{3}]-doublet template of \cite{Sergeev2007}, 
in which the doublet lines was assumed to have the same profile with an intensity ratio of 2.93.  
Following the approach described in \cite{Sergeev2007}, we first remove the continuum
by subtracting a straight line interpolated between two 40 \AA-wide windows centered 
on 4640~\AA\ and 5110~\AA\ in the rest frame. We then remove the [\ion{O}{3}] 
doublet using the template and remove the narrow H$\beta$ component using the same template but 
adjusting its position and intensity to achieve the smoothest broad-line H$\beta$ profile.
The obtained flux ratio of narrow H$\beta$ to [\ion{O}{3}]~$\lambda5007$ generally 
ranges from 0.10 to 0.15, consistent with the previous determinations 
(\citealt{Peterson2004, Sergeev2007}). After obtaining the broad H$\beta$ line, 
its integrated flux is calculated over a uniform wavelength window 4714--4933~{\AA} in
the rest frame, as in \cite{Peterson2002}.

\begin{figure*}[t!]
\centering
\includegraphics[angle=0,width=1.0\textwidth]{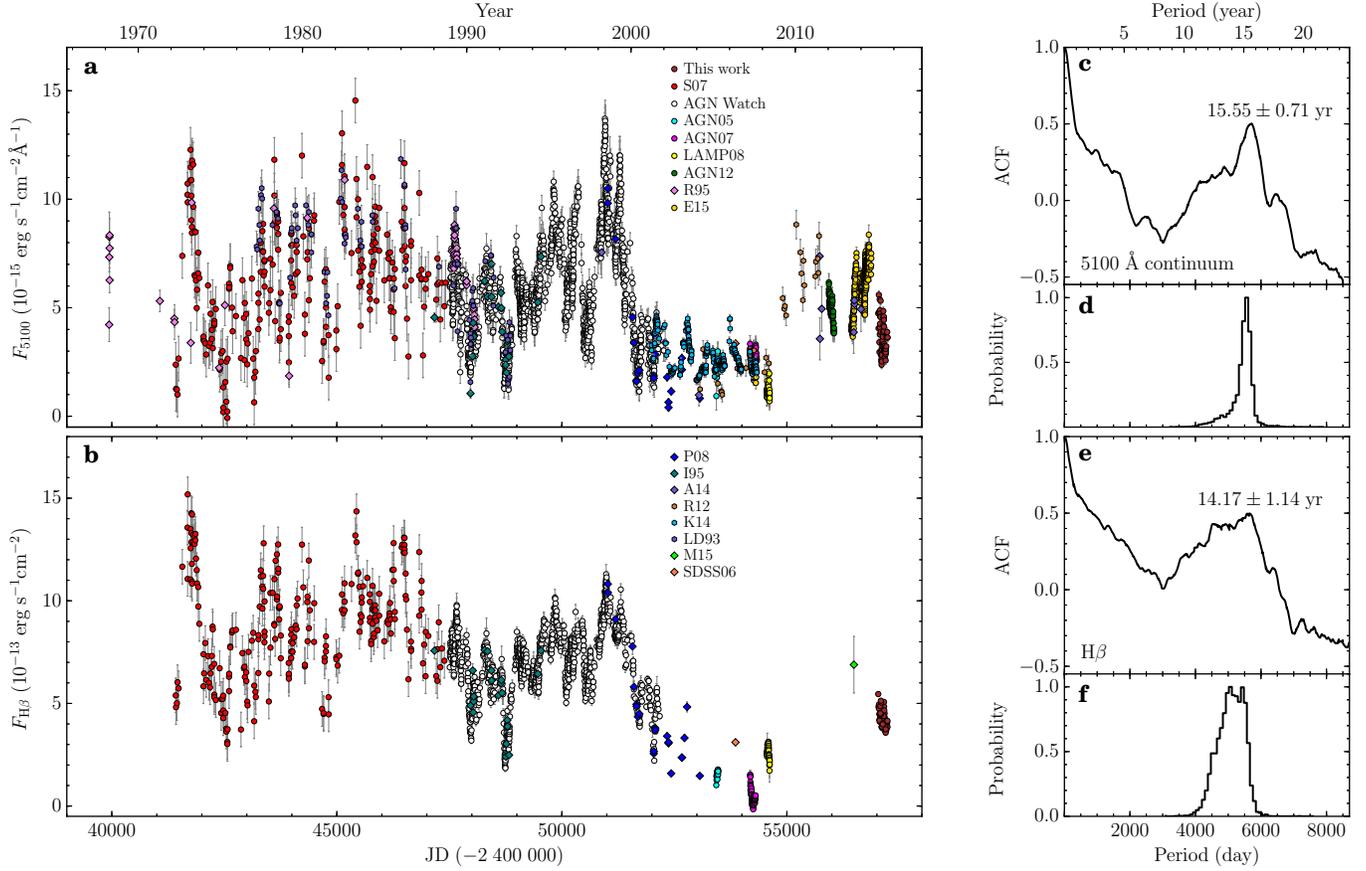}
\caption{\footnotesize
Period determination from the light curves in ({\it a}) 5100~\AA\ continuum  and 
({\it b}) H$\beta$. Data sources are labelled and explained in Appendix A and 
Tables \ref{tab_hb} and \ref{tab_con}.  The ACF and distribution of the peak period for 
the continuum are shown in panels ({\it c}) and ({\it d}), and for H$\beta$ in panels ({\it e}) 
and ({\it f}).}
\label{fig_lc}
\end{figure*}

\begin{figure*}[t!]
\centering
\includegraphics[angle=0,width=0.5\textwidth]{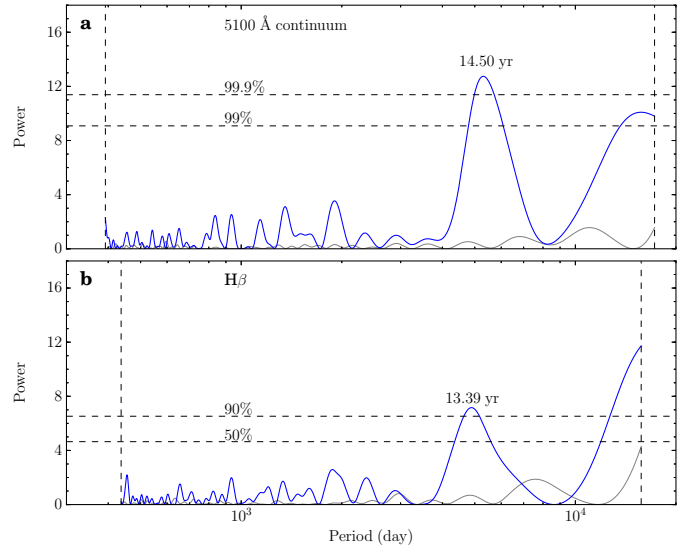}
\caption{\footnotesize
Lomb$-$Scargle periodograms of the light curves in ({\it a}) 5100~{\AA} continuum and 
({\it b}) H$\beta$ fluxes (blue lines), using the procedure given by \cite{Press1992}.
Solid grey lines show the periodograms of the data window, which help uncover spurious signals caused 
by the sampling; none is seen. Dashed vertical lines represent the Nyquist frequency and 
the time span of the light curve, respectively, and dashed horizontal lines represent 
the confidence levels. }
\label{fig_lsp}
\end{figure*}

It is worth stressing that in the decomposition procedure described above, we ignore
\ion{Fe}{2} blends and the broad \ion{He}{2}~$\lambda4686$ emission line. In addition, 
the absorption of host galaxy at H$\beta$ line will also deform the obtained H$\beta$ profiles.
In Appendix~B, we perform a simple test to show that host galaxy contamination 
is unimportant on the H$\beta$ profiles when the flux ratio 
between the central AGN and the host galaxy at 5100~{\AA} is larger than $F_{\rm AGN}/F_{\rm gal} \approx 0.5$; 
otherwise, the host galaxy contamination leads to apparent double peaks in the 
H$\beta$ profile. The resulting changes in the integrated H$\beta$ fluxes are less 
than 10\% for $F_{\rm AGN}/F_{\rm gal}\gtrsim0.2$, which is generally satisfied in our 
database (see Figure~\ref{fig_lc}). Therefore, we use all the data for analyzing the 
light curves, but exclude those spectra with small $F_{\rm AGN}/F_{\rm gal}$ for analyzing 
H$\beta$ profiles. We adopt a lower limit of $F_{\rm AGN}/F_{\rm gal}\approx0.7$ to be conservative. 
The spectroscopic data sources AGN05, AGN07, and LAMP08, together with several spectra 
in other data sources---in total 85 out of 924 spectra---are finally discarded. 

\ion{Fe}{2} emission is relatively weak in NGC~5548 (e.g., \citealt{Wamsteker1990, 
Vestergaard2005, Denney2009, Mehdipour2015}). \cite{Vestergaard2005} selected 73 
spectra of NGC~5548 spanning 13 years of the AGN Watch project and 
quantitatively showed that, on average, the flux ratio of optical \ion{Fe}{2} over 
the range 4250--5710~{\AA} (rest-frame) to broad H$\beta$ line is 0.66.  The 
strength of \ion{Fe}{2} emission in the 
H$\beta$ window is fairly small and is 
thus expected to have little effect on the broad H$\beta$ profile.

The broad \ion{He}{2}~$\lambda4686$ line blends with the blue wing of 
H$\beta$ (\citealt{Bottorff2002}). \cite{Denney2009} studied exhaustively 
the contamination of \ion{He}{2}~$\lambda4686$ on H$\beta$ for NGC~5548, again
using the spectra from the AGN Watch project. Their results showed that 
although deblending \ion{He}{2} slightly increases the flux of 
the far blue wing of H$\beta$, its influence on the H$\beta$ line width (in term of full 
width at half maximum) is safely negligible (see their Figures~16 and 19). We 
also expect contamination from \ion{He}{2}~$\lambda4686$ to be unimportant. 

In Figure~\ref{fig_example}, we show three illustrative examples for our 
spectral decomposition, with the first two spectra from the AGN Watch project, taken during 
high and low state, respectively, and the other one from our new observations.
As can be seen, it is safe to ignore the \ion{Fe}{2} emission
and the broad \ion{He}{2} line for our present purposes.

\section{Periodicity of long-term variations}

The optical photometric and spectroscopic data cover a time span of over four decades. 
Figures \ref{fig_lc}{\it a} and \ref{fig_lc}{\it b} plot light curves of the 
5100~\AA\ continuum and H$\beta$ fluxes, respectively. Appendix~A shows all 
the H$\beta$ profiles. It is obvious at first glance that light curves of both
the 5100~\AA\ continuum and H$\beta$ appear periodic. Below we first 
summarize the properties of the long-term variations in the continuum, H$\beta$ flux, 
and as well as H$\beta$ profile, and then explore the periodicity in these variations.

\subsection{Properties of long-term variations}

The continuum and H$\beta$ light curves of NGC~5548 exhibit three types of 
variations: 

\begin{enumerate}
\item{
Short-term fluctuations of a factor of $1.5-2$ at any epoch, which could 
be driven by local instabilities of the accretion disk(s).  They are well-modeled 
as a damped random walk (\citealt{Kelly2009,Li2013}), with a characteristic 
timescale that depends on the optical luminosity.  The average timescale over 
16 monitoring campaigns is about $\tau_c\approx 150$ days (\citealt{Li2013}). 
This kind of variations is used for measurements of H$\beta$ reverberation 
mapping to reveal the BLR structure and dynamics. The H$\beta$ flux variations 
lag from 5 to 30 days behind the continuum variations, depending on the continuum flux 
state (e.g., \citealt{Bentz2013}).
}

\item{Long-term variability by a factor of $\sim 10$ on a timescale of 10 
yr, which, according to our analysis, appears to be periodic and governed 
by the orbital motion of a SMBHB. This type of variability has the largest 
amplitude.}

\item{Abrupt changes by a factor up to $3-4$, such as the large dip in the 
5100~\AA\ continuum in 1981$-$1982, in 1996, and in 1997$-$1998, and the strong 
flare in 1972$-$1973 and 1998--1999.  The causes for these irregular changes 
are unknown. Such patterns of abrupt changes is very similar to that seen in 
OJ~287, which has long been suspected to host a SMBHB system 
(e.g., \citealt{Sundelius1997,Valtonen2008}).}
\end{enumerate}

The broad H$\beta$ line of NGC~5548 generally always has a complex, asymmetric 
profile (e.g., \citealt{Shapovalova2004}), which has been suggested to be caused 
by a superposition of several physically distinct components that vary with time 
(e.g., \citealt{Peterson1987,Sergeev2007}).  From the four decades of record,
we can generally identify four types of H$\beta$ profiles:
\begin{enumerate}
\item An asymmetric shape during some high states. 

\item A very broad and highly asymmetric shape with a width of $\sim10^4\,\kms$ 
during low states.

\item A double-peaked shape during some intermediate-luminosity states.

\item A highly asymmetric, shape dominated by a blue component seen around 1972 and 1985.

\end{enumerate}
As we will show below, the profile variations are consistent with orbital 
motion of two BLRs associated with a SMBHB system.  The period inferred from 
the profile changes agrees with the periods observed in the 5100\,\AA\, 
continuum and H$\beta$ flux series.

\begin{figure*}[t!]
\centering
\includegraphics[angle=0,width=0.7\textwidth]{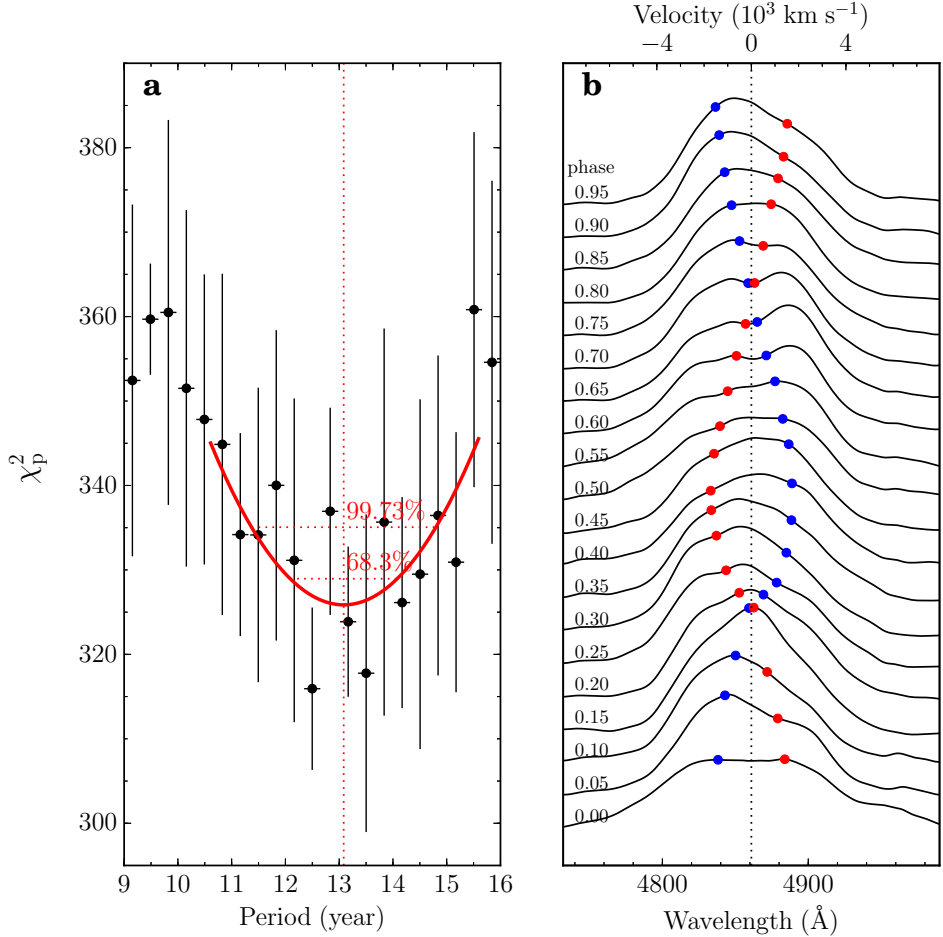}
\caption{\footnotesize 
({\it a}) $\chi_{\rm p}^2$ versus period from the epoch-folding scheme to determine the period 
($T_{P_{\rm H\beta}}$) of H$\beta$ profile variations, in which the phase width is set  to 
$\Delta \phi=150$\,days (see Section 3.3 for details). We fit the 
$\chi_{\rm p}^2-T_{P_{\rm H\beta}}$ relation with a second-order polynomial and 
determine the period by the minimum of the polynomial, which yields $\TPhb=(13.10\pm 0.88)$\,yr.  
The uncertainty is determined by $\Delta \chi_{\rm p}^2=3.1$, which corresponds 
to a confidence level of $68.3\%$ (``$1\sigma$''). ({\it b}) Folded H$\beta$ profile series 
with phases for a complete cycle using the mean period 14.14 yr. To guide the eye, 
blue and red points are superposed to show the line-of-sight velocities of the black hole pair
from our SMBHB model (see Section 5).
}
\label{fig_phase}
\end{figure*}

%
\begin{deluxetable*}{clcl}
 \tablecolumns{4}
 \tablewidth{0.8\textwidth}
 \tablecaption{Summary on the observed periods of 5100~{\AA} continuum flux, H$\beta$ flux, 
               and H$\beta$ profile in NGC~5548}
 \tabletypesize{\footnotesize}
 \tablehead{
 \colhead{Period}       &
 \colhead{Meaning}      &
 \colhead{Value (yr)}  &
 \colhead{Method}
 }
\startdata
$T_{\rm 5100}$        & Period of 5100~{\AA} continuum        & $15.55\pm0.71$   & ACF method\\
                      &                                       & $14.50$          & Lomb-Scargle periodogram\\
$\TFhb$               & Period of integrated H$\beta$ fluxes  & $14.17\pm1.14$   & ACF method\\
                      &                                       & $13.39$          & Lomb-Scargle periodogram \\
$\TPhb$               & Period of H$\beta$ profiles           & $13.10\pm0.88$   & Epoch-folding scheme\\
Mean                  & \nodata                               & $14.14\pm0.87$          & \nodata
\enddata
\tablecomments{The uncertainty of the mean period is set by the standard deviation of the five observed periods.}
\label{tab_period}
\end{deluxetable*}


\begin{figure}[t!]
\centering
\includegraphics[angle=0,width=0.4\textwidth]{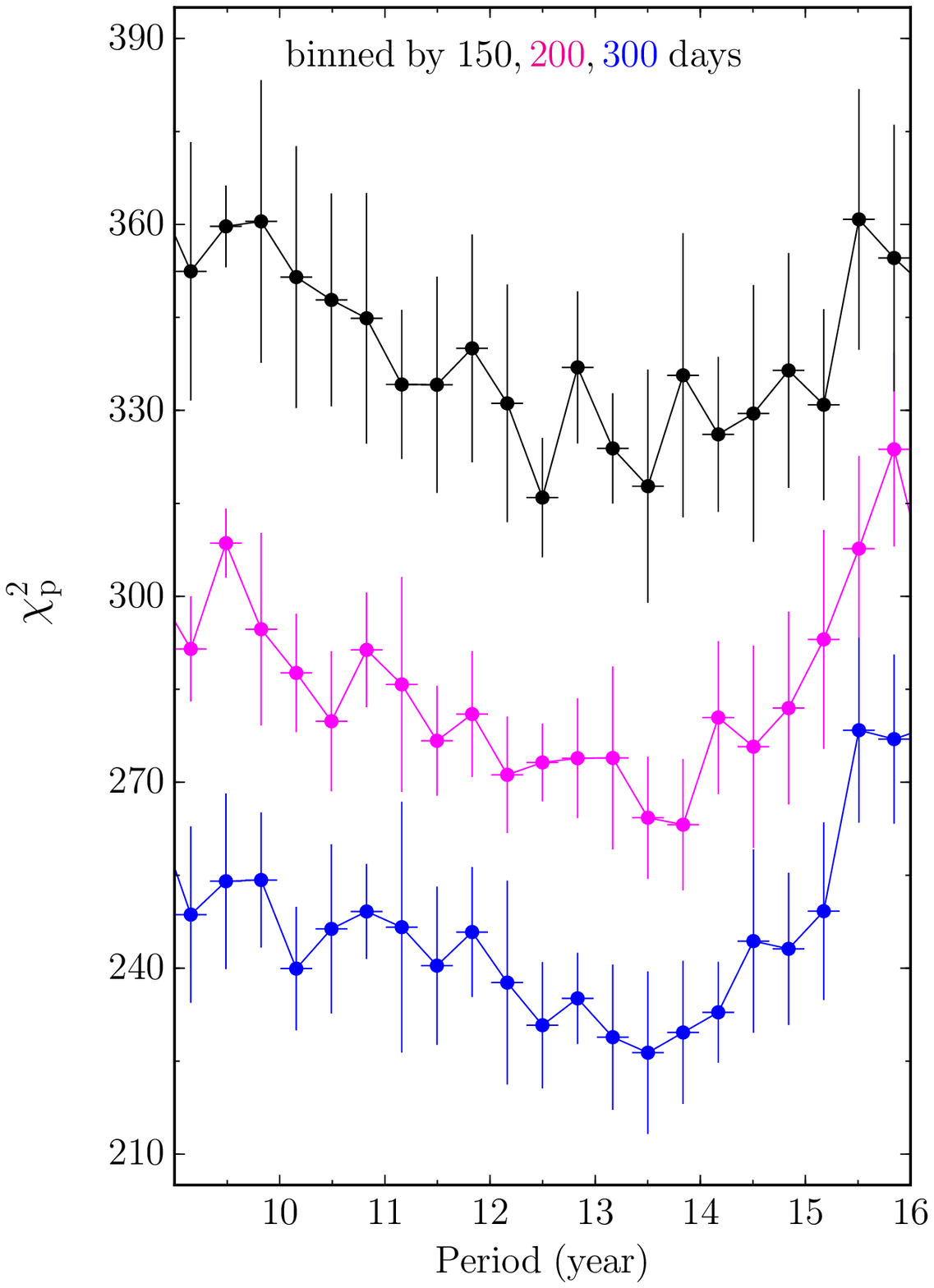}
\caption{\footnotesize 
$\chi_{\rm p}^2$ versus period for different phase widths ($\Delta\phi=150,200,300$\,days) of the 
epoch-folding scheme (see Section 3.3 for details). As can be seen, the period of H$\beta$ profile 
variations $\TPhb$, determined from the minimum $\chi_{\rm p}^2$, is insensitive to the choice 
of $\Delta \phi$. 
} 
\label{fig_phase2}
\end{figure}

\subsection{5100~\AA\ continuum and H$\beta$ line fluxes}
We applied the auto-correlation function (ACF) technique to identify the periods 
of continuum and emission-line flux variations. We use the interpolation 
cross-correlation function method  of \cite{Gaskell1987} to calculate the ACFs and 
the model-independent FR/RSS Monte Carlo method described by \cite{Peterson1998} to 
determine the associated uncertainties. The period is determined by the centroid 
of the ACF around its peak (excluding the zero peak) above a threshold of 80\% of 
the peak value (\citealt{Peterson1998}). To suppress the short-term fluctuations 
in the light curves (see Section 3.1), we bin the light curves with an interval of 
150 days, the characteristic timescale of fluctuations. The uncertainties of 
points in the newly binned series are assigned as follows: we first calculate 
the mean standard deviation of the whole light curve, and then compare it with 
the standard deviations of each bin and set the associated uncertainty to be the 
maximum one. Such a choice can account for the sparse sampling in some periods
(e.g., around 2010) and avoid unrealistic weights of these measurements in 
calculating the ACFs.

Figures \ref{fig_lc}{\it c}$-$\ref{fig_lc}{\it f} show the ACF results and 
period distributions from standard Monte Carlo simulations (\citealt{Peterson1998}). 
We detect highly significant peaks in the ACFs of the 5100~\AA\ continuum and H$\beta$ 
light curves that correspond to periods of $T_{5100}=15.55\pm 0.71$ yr 
and $\TFhb=14.17\pm1.14$ yr, respectively, where the uncertainties are 
determined by the period distributions with a 68.3\% confidence level (``1$\sigma$'').  
We note that $\TFhb\approx T_{5100}$, which is likely a direct consequence of photoionization of 
the BLR by the central continuum.  

We also apply the Lomb-Scargle normalized periodogram (e.g., \citealt{Press1992})
to test for periodicity in the light curves. Again, the light curves are binned 
with an interval of 150 days. Figure~\ref{fig_lsp} shows the Lomb-Scargle
periodograms over a period range limited by the average Nyquist frequency and the time 
span of the light curve, using the procedure given by \cite{Press1992}
\footnote{We set the parameter $\texttt{hifac}=1$ to impose a lower limit on the period in 
calculating the Lomb-Scargle periodogram by the Nyquist frequency.}. 
The  periodogram of the 5100~{\AA} continuum peaks at 14.50 yr with a confidence level 
$>99.9$\%; for the H$\beta$ fluxes, it peaks at 13.39 yr with a confidence level $>90$\%.
Here, the confidence level represents the probability that the data contain a
periodicity and can be tested by the statistic $1-p$, where $p$ is the false alarm 
probability of the null hypothesis (\citealt{Press1992}). To exclude spurious 
signals caused by the sampling, we also plot the periodograms of data window
{\footnote{See \citet[p.92]{Hilditch2001} for the definition of the Lomb-Scargle 
periodogram of data window.}} 
in Figure~{\ref{fig_lsp}. As can be seen, the periodicity cannot arise from 
the sampling. The obtained periods are generally consistent with those from 
the ACF method.

Although there exists alternative explanations, a periodic variability in time 
series of flux measurements is widely used to search for SMBHB candidates 
(e.g., \citealt{Graham2015, Graham2015b, Liu2015}). The best-known candidate is 
OJ~287, which exhibits outburst peaks every $\sim12$ years (\citealt{Valtonen2008}). 
Recently, \cite{Graham2015b} performed a systematic search for periodic light curves in 
quasars covered by the Catalina Real-time Transient Survey and obtained a sample of more 
than 100 candidates. However, as the authors note, more spectroscopic 
observations are required to verify the SMBHB hypothesis. Nevertheless, the periodicity 
of the H$\beta$ and continuum long-term variations suggests that NGC~5548 may host 
a SMBHB system. Compared with the above candidates, NGC~5548, at a distance of 
$\sim75$ Mpc, is to date one of the nearest objects with periodic behavior. 
Another nearby case is NGC~4151 at $\sim 14$ Mpc ($z=0.003319$), which, interestingly, was also found to 
have a similar period of $\sim15$ yr in flux variations (\citealt{Pacholczyk1983, Guo2003}). 
The SMBHB hypothesis in NGC~4151 was subsequently reported by \cite{Bon2012} based on an 
analysis of the H$\alpha$ emission line. Below we use the extensive spectroscopic 
monitoring database over four decades (much longer than the time span of the database 
for NGC~4151) to provide more evidence for a SMBHB in NGC~5548.

\subsection{H$\beta$ profiles and epoch-folding scheme}
The H$\beta$ profiles of NGC~5548 generally show two peaks superposed on a 
very broad wing, however, during most epochs the double peaks are indistinct and not 
well-separated{\footnote{Indeed, it was already pointed out by \cite{Chen1989}
that, simply from the virial relation, the projected velocity separation between 
the two BLRs of a SMBHB should be less 
than their individual line widths (see also 
\citealt{Shen2010, LiuJ2015}). Therefore, the double peaks are expected to be highly blended,
and widely separated peaks generally are not the signature of BLRs bound to a SMBHB.}. 
This prevents us from unambiguously testing the SMBHB model as in previous works by 
directly comparing the velocity changes of the double peaks with the orbital motion 
of a SMBHB model (e.g., \citealt{Eracleous1997}). Notwithstanding, over time the 
H$\beta$ profiles clearly exhibit systematic variations: the peaks shift and vary, 
plausibly periodically, on a long timescale (see the series of spectra over 
four decades in Appendix A).  As illustrated in Appendix B, the variations cannot 
be due to host galaxy contamination.

H$\beta$ has been monitored over more than two cycles of the orbital period of 
$\sim 14$ yr, long enough to use the epoch-folding scheme to test for periodicity 
in the line profile variations. Our epoch-folding scheme is as follows. 
Supposing a trial period $\TPhb$, we group the temporal series of the normalized 
H$\beta$ profiles into $\left\{\left[\left(t_N-t_0\right)/\TPhb\right] +1\right\}$ 
uniformly spaced cycles, where the brackets of $[B]$ indicate the integral 
part of number $B$, and $t_0$ and $t_N$ are the starting and ending 
times of the series, respectively.  In each cycle, we divide the series into 
$N_\phi=\TPhb/\Delta\phi$ uniformly spaced phases by a phase width 
$\Delta\phi$. For an H$\beta$ spectrum at time $t_j$, it will be grouped into 
the phase of $\left\{(t_j-t_0)/\TPhb-[(t_j-t_0)/\TPhb]\right\}$ in the cycle of 
$\left\{\left[\left(t_j-t_0\right)/\TPhb\right]+1\right\}$.  In the $k$th-phase 
of the $i$th-cycle, we calculate the averaged spectrum $\bar{F}_{\lambda}^{i,k}$ 
and the standard deviation $\sigma_{\lambda}^{i,k}$, and then calculate the averaged 
spectrum $\widehat{\bar{F}}_{\lambda}^{k}$ of the $k$th-phase over all the cycles.  
Finally, we compare the differences of the averaged spectra in the same phase of 
different cycles by defining a statistical quantity
\begin{equation}
\chi_{\rm p}^2(\TPhb)=\frac{1}{N_{\phi}}\sum_k\frac{1}{N_k-1}\sum_{N_{\lambda}}
   \sum_{i}\left(\bar{F}_{\lambda}^{i,k}-\widehat{\bar{F}}_{\lambda}^{k}\right)^2
   \Bigg/\left(\sigma_{\lambda}^{i,k}\right)^{2},
\end{equation}
where $N_{\lambda}$ is the wavelength number of H$\beta$ profiles and 
$N_k$ is the number of cycles for the $k$th-phase.  The minimum of 
$\chi_{\rm p}^2$ means that the H$\beta$ lines with a time difference $\TPhb$
have systematically similar profiles. In other words, the H$\beta$ profiles 
have periodic variations with a period $\TPhb$. It is worth stressing the two following
points: (1) the H$\beta$ profiles are normalized to avoid the influence of the 
periodicity in the integrated H$\beta$ fluxes; (2) the random fluctuations 
of the profiles in the $k$th-phase are suppressed by averaging the spectra in 
the calculation of $\chi_{\rm p}^2$.

For the datasets of NGC~5548, the wavelength number is large ($N_\lambda\gg1$).
It can be shown that the quantity $\chi_{\rm p}^2$ approximately follows a normal
distribution with a mean of $N_\lambda$ and a standard deviation of 
$\sqrt{2N_\lambda/(N_k-1)N_\phi}$. We determine the confidence interval of $\TPhb$ by 
$\Delta \chi_{\rm p}^2 = 3.1$, which corresponds to a 68.3\% confidence level (``$1\sigma$'').
In our calculations, we adopt the phase width $\Delta\phi=150$ days. We use a second-order
polynomial to fit the $\chi_{\rm p}^2-\TPhb$ relation and determine the period 
by the minimum of the polynomial. Figure \ref{fig_phase}{\it a} shows that $\chi_{\rm p}^2$ 
reaches a minimum at $\TPhb=(13.10\pm 0.88)$ yr, which indicates the period of H$\beta$ profile 
variations. In Figure \ref{fig_phase2}, we test the dependence of $\chi_{\rm p}^2$ 
on the phase width $\Delta\phi$ and find that the resulting periods from the 
minimum of $\chi_{\rm p}^2$ are insensitive to the choice of $\Delta\phi$. 

Since the 5100~{\AA} continuum of NGC~5548 varies periodically, we 
expect that the line width of broad H$\beta$ line should also follow such a
temporal variation pattern ($\sigma({\rm H\beta})\propto F_{\rm con}^{-1/4}$), 
simply from the well-established relationship between the BLR size and AGN 
5100~{\AA} luminosity (the $R_{\rm BLR}-L$ relation; e.g., \citealt{Bentz2013}) 
and the virial relation of the BLR motion (e.g., \citealt{Peterson1999, 
Peterson2004, Simic2016}).
This will contribute to the periodicity of H$\beta$ profiles seen in 
the above epoch-folding scheme. In Appendix~C, we perform a simple simulation test
to show that the periodicity in variations of the H$\beta$ profiles does not 
entirely result from such a dependence of the H$\beta$ line width on the 
continuum.  

To clarify long-term variations of the two peaks, we folded the spectra over 
four decades in light of phases using the mean period 14.14 yr (see Table~\ref{tab_period}). 
Figure \ref{fig_phase}{\it b} shows the folded spectra series with phases for one 
complete cycle. The two peaks of H$\beta$ profiles seem to shift in an 
opposite direction. To guide the eye, we also superpose the line-of-sight velocities of 
the black hole pair (red and blue points) inferred from our SMBHB BLR model, as described below.

Within uncertainties, the three periods of variability, namely that of the 
H$\beta$ profile, H$\beta$ flux, and continuum flux, are all in agreement, i.e., 
$\TPhb\approx \TFhb\approx T_{5100}$. 
Combining with the tidal imprints of a major merger event in the morphology of
the host galaxy, this constitutes very compelling spectroscopic evidence
for the presence of a SMBHB in NGC~5548. 

NGC~5548 contains a classical bulge (\citealt{Ho2014}) with a central stellar velocity 
dispersion of $\sigma_*=195\pm 13\,\kms$ (\citealt{Woo2010}).  This yields a (total) black 
hole mass of $\log\left(\bhm/\sunm\right)=8.44\pm0.14$ from the $\bhm-\sigma_*$ relation 
(\citealt{Kormendy2013})%
{\footnote{We note that the H$\beta$
reverberation mapping campaigns of NGC~5548 yield a black mass
of $\sim 8\times10^7M_\odot$ (e.g., \citealt{Bentz2009, Lu2016}; see also \citealt{Ho2015}), 
generally smaller than the value from the $M_\bullet-\sigma_\star$ relation. Considering that 
the reverberation-mapping mass relies on a so-called 
virial factor, which is calibrated by the $M_\bullet-\sigma_\star$ relation, we prefer to
use the direct estimate from the $M_\bullet-\sigma_\star$ relation. However, 
a smaller black hole mass does not qualitatively affect our results.}}.
Using this mass as the total mass of the SMBHB system and 
the mean observed period as the (redshifted) orbit period (Table~\ref{tab_period}), 
we determine the semi-major axis of the orbit from Kepler's third law, $\bha=21.73\pm 2.43$ 
light-days (Table~{\ref{tab_par}).

\section{Confronting alternative interpretations}
Besides the SMBHB model, there are several alternative models/interpretations
 for the periodicity in the light curves and line profiles; 
however, most cannot simultaneously 
account for all of the evidence in NGC~5548.  Double-peaked profiles of broad 
emission lines quite often appear in low-luminosity AGNs (\citealt{Eracleous1995, 
Ho2000, Strateva2003, Lewis2010}), and can be understood in the context of the 
radiatively inefficient accretion flow structure of AGNs accreting at highly 
sub-Eddington rates (\citealt{Ho2008}). NGC~5548 has an Eddington ratio 
$\gtrsim 10^{-2}$ (\citealt{Ho2014}); it does not belong to the class of 
low-accretion rate systems.  The central (or binary) black hole mass of NGC~5548, 
$\bhm\approx10^8\sunm$, is quite secure, both from reverberation mapping and from 
the bulge stellar velocity dispersion (\citealt{Ho2015}).  Reverberation mapping 
finds the H$\beta$ BLR radius lies in the range $5-30$ light-days (\citealt{Bentz2013}), 
corresponding to $\rhbeta\sim10^3R_{\rm g}$, where $R_{\rm g}=G\bhm/c^2$ is the 
gravitational radius.   These facts, together with the periodicity, allow us to 
evaluate the interpretations listed below.

\subsection{Disk model}
Double-peaked broad emission lines are commonly modeled as disk emission. 
We dismiss the disk model for NGC~5548 based on the following arguments. A
simple circular disk model can be easily ruled out because, due to
relativistic boosting, the blue peak should always be stronger than the red 
peak, clearly contrary to what is observed.  In the precessing eccentric disk 
model (\citealt{Eracleous1995}), the precession period due to the relativistic 
advance of the pericenter can be expressed as (\citealt{Weinberg1972})
\begin{equation}
T_{\rm prec}=\frac{2\pi}{3}\frac{1+e}{(1-e)^{3/2}}\frac{G\bhm}{c^3}R^{5/2}
=10^3\frac{1+e}{(1-e)^{3/2}}m_8r_3^{5/2}\,{\rm yr},
\end{equation}
where $e$ is the eccentricity, $m_8=\bhm/10^8\,\sunm$, and 
$r_3=\rhbeta/10^3\,R_{\rm g}$. The term $(1+e)/(1-e)^{5/2}$ is always larger 
than~1, and as a result the precession period is $>10^3$ yr, making this 
model implausible.

For the case of precessing spiral arms in a disk (\citealt{Storchi-Bergmann2003}),
an upper limit bound to the pattern period is set by the sound-crossing 
timescale, 
$T_{\rm s}=R/c_{\rm s}=66m_8r_3T_5^{-1/2}\,{\rm yr}$,
where the disk temperature $T_5=T/10^5$ K, beyond which thermal dissipation 
will smear out the spiral arms.  Numerical simulations by \cite{Laughlin1996}
show that the pattern speed is several times to an order of magnitude larger 
than the dynamical timescale,
$T_{\rm dyn}\sim(R^3/GM_\bullet)^{1/2}=0.5m_8r_3^{3/2}$~yr.
The pattern period is between $T_{\rm dyn}$ and $T_{\rm s}$, which in general 
can match the observed period. Given the many unconstrained free parameters 
in the spiral-arm disk model, with sufficient tuning it certainly  should be 
able to reproduce the observed H$\beta$ profiles.  However, this model takes no account of
the periodic variations in 5100~{\AA} continuum and H$\beta$ fluxes.

It is possible the thermal instability of the accretion disk leads
to repeat outbursts of accretion onto the black hole and gives rise to periodic 
continuum variations, but this generally occurs on a much long timescale of 
the order of $10^3-10^5$~yr (e.g., \citealt{Mineshige1990,Hameury2009,
Janiuk2011}). Even if the instability were to develop only in a part of the 
disk and the continuum variations are modulated by the disk rotation, 
the amount of the modulation (i.e., the Doppler boosting) in  
the regions that produce the 5100~{\AA} continuum (at $\sim10^3 R_{\rm g}$ with 
a period of $\sim14$~yr) is insufficient for the 
observed large amplitude (see also Section 4.3). 
Another possibility is that the spiral
arms may trigger periodic accretion onto the central black hole and thereby
may result in the periodic variability in the continuum (Y. Shen, private communications). 
\cite{Chakrabarti1993} studied the variations in emission from accretion disks with spiral shocks and 
tended to use such a model to explain the micro-variability in AGNs.
It is unclear yet if spiral arms can drive the large-amplitude variations 
observed in NGC~5548.

\subsection{Precessing jet model}

The expected precession period for a single SMBH is given by 
\begin{equation}
P=10^{9.25}\alpha^{48/35}a^{5/7}m_8^{1/7}
   \left(\frac{\dot{M}_{\bullet}}{10^{-2}~M_\odot{\rm yr}^{-1}}\right)^{-6/5}~{\rm yr},
\end{equation}
where $\alpha$ is the viscosity parameter, $a$ is black hole spin, and 
$\dot{M}_{\bullet}$ is the mass accretion rate (\citealt{Lu2005}). This period is typically much 
longer than the observed period unless $a$ approaches extraordinarily close 
to 0. On the other hand, the radio-loudness parameter of NGC~5548's nucleus 
is $\log {\cal R}=1.24$ (\citealt{Ho2001}), meaning that it is only marginally 
radio-loud. The global spectral energy distribution for NGC~5548 also suggests
that the 5100~{\AA} continuum predominantly stems from an accretion disk 
(\citealt{Chiang2003, Mehdipour2015}).  We can thus rule out with certainty
that the periodicity originates from a precessing jet.

\subsection{Hot spot model}
In this model, a bright hot spot embedded in the accretion disk orbits around the 
central black hole and produces an excess, time-varying component in line emission
(e.g., \citealt{Newman1997, Jovanovi2010}). There are no physical constraints for 
most of the free parameters of the model (e.g., amplitude, width, and location of the spot, etc.).
Fine-tuning some of them, one can generally simulate complicated H$\beta$ profiles.
However, as in the disk models, this model does not account for
the periodic variations in 5100~{\AA} continuum and H$\beta$ fluxes. From the 
reverberation-mapping observations of the H$\beta$ line (\citealt{Bentz2013}), 
we know that the orbit size of the hot spot is $R_{\rm spot}\approx10^3R_{\rm g}$, 
which is located at the outer region of the disk. Assuming a Keplerian orbit, 
the velocity of the spot is $\beta=V_{\rm spot}/c\approx0.03\,(R_{\rm spot}/10^3R_{\rm g})^{-1/2}$.
The resulting Doppler boosting factor 
\begin{equation}
\delta=\frac{1}{\varGamma(1-\beta\cos\varphi\cos i)}\approx 1 + \beta \cos\varphi\cos i,
\end{equation}
is confined to a range $1\pm0.03$, where $\varGamma=(1-\beta^2)^{-1/2}$ is 
the Lorentz factor, $i$ is the orbital inclination angle, and $\varphi$ is the
phase angle. If this model works in NGC~5548, the amount of 
Doppler beaming is too small to produce the observed variability amplitude 
of the integrated H$\beta$ fluxes. On the other hand,   
the disk instability predicted by existing models also cannot appropriately 
account for the large amplitude variation of the 5100~{\AA} continuum (Section 4.1). 
Taken together, these facts do not support the hot spot model.

\subsection{Warped accretion disk model}

A warped accretion disk will precess around the central black hole and 
eclipse some parts of the continuum  (e.g., \citealt{Pringle1996}), 
leading to periodic variations in the continuum. 
There are two types of warps in disks surrounding single SMBHs: 
irradiation-induced warps by a central radiation source (\citealt{Pringle1996}), and 
Lense-Thirring torque-induced warps by a spinning black hole (e.g., \citealt{Bardeen1975, Martin2007}). 

For the irradiation-induced warps, the condition of warp 
instability is 
\begin{equation}
\frac{R_{\rm warp}}{R_{\rm g}}>16\pi^2\eta^2\epsilon^{-2},
\end{equation}
where $\eta$ is the ratio of the viscosity in the vertical and azimuthal 
directions and $\epsilon$ is the accretion efficiency. This gives 
$R_{\rm warp}\sim10^4R_{\rm g}$ if we set $\eta\approx1$ and 
$\epsilon\approx0.1$. The precession timescale is 
\begin{equation}
P_{\rm prec}=\frac{12\pi c \Sigma R_{\rm warp}^3\Omega_{\rm Kep}}{L}
            \approx\frac{12\pi c M_{\rm disk}R_{\rm warp}\Omega_{\rm Kep}}{L},
\end{equation}
where $c$ is the speed of light, $\Sigma$ is the surface density, 
$\Omega_{\rm Kep}$ is the orbital angular velocity, and $M_{\rm disk}$ 
is the mass of the warped disk (\citealt{Pringle1996, Storchi-Bergmann1997}).  If we adopt 
$M_{\rm disk}\sim(10^{-2}-10^{-3})\bhm$ and $L=10^{43}\,{\rm erg~s^{-1}}$, the period 
is on the order of $10^{6-7}$ yr, much longer than the observed period of 
$\sim14$ yr. 

The Lense-Thirring torque-induced warps occur when a disk is inclined with respect to the 
spin axis of the central black hole (e.g., \citealt{Martin2007}). The warp radius 
is typically $300R_{\rm g}$, depending on the black hole spin, and the precessing timescale 
when only considering the Lense-Thirring torque is on the order of $10^{4-5}$ yr (\citealt{Li2013a, Li2015}).
Recently, \cite{Ulubay2009} numerically studied the precession of warped disks 
when the self-gravity of the warped disk dominates the angular-momentum precession rate
(the so-called self-gravitating warped disks; see also \citealt{Tremaine2014}).
The precession rate is found to be proportional to the mass ratio of the disk to the black hole as
\begin{equation}
\dot\phi \sim f\frac{M_{\rm disk}}{M_\bullet}\Omega_{\rm Kep} 
= f \frac{M_{\rm disk}}{M_\bullet}\frac{c}{R_{\rm g}}\left(\frac{R_{\rm g}}{R_{\rm warp}}\right)^{3/2},
\end{equation}
where $f$ is a scalar coefficient and lies in a range of $10^{-2}-1$ (see Figure~9
of \citealt{Ulubay2009}). This results in a precession period of
\begin{equation}
P_{\rm prec} = \frac{2\pi}{\dot \phi}\sim 50-5\times10^4~{\rm yr},
\end{equation}
where we adopt $R_{\rm warp}=300R_{\rm g}$ and 
estimate $M_{\rm disk}\approx hM_\bullet\sim(10^{-3}-10^{-2})M_\bullet$,
beyond which the disks will be gravitationally unstable (e.g., \citealt{King2008}).
Here $h$ is the disk's aspect ratio. The lower limit is marginally larger 
than the observed period of 14 yr, implying that a warped
disk model might explain the periodicity in the 5100~{\AA} continuum
in NGC~5548.
However, it is unknown if the warps can simultaneously eclipse some parts
of the BLR, so as to produce the periodically varying H$\beta$ profiles.
In addition, the light curves from a precessing warped accretion disk are 
expected to be sinusoidal, inconsistent with the observed variation pattern. 
In a nutshell, we cannot firmly rule out the warped disk model, 
but in many aspects it seems implausible.

\begin{figure}[t!]
\centering
\includegraphics[angle=-90,width=0.5\textwidth]{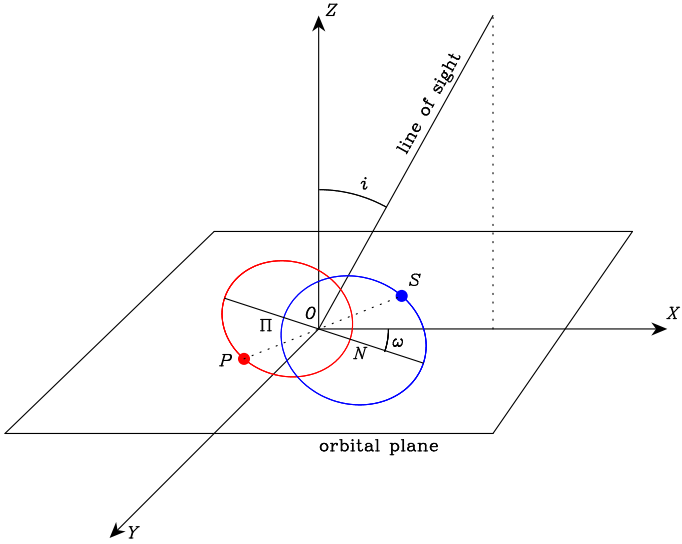}
\caption{\footnotesize 
Schematic for the geometry of the SMBHB system. The orbital plane is 
located in the $XOY$-plane, and the observer is located in the $XOZ$-plane. 
$\Pi$ and $N$ denote locations of the periastron.  $P$ and $S$ denote 
locations of the primary and secondary black hole at the start of 
observations ($t=0$). The parameter $\omega$ denotes the angle of the 
periastron with respect to the $X$-axis.  The parameter $\phi_0$ denotes the 
fraction of an orbit that the periastron occurs, prior to $t=0$.  Thus, at 
$t=0$, the primary black hole is at an angle $\protect\angle NOP$ from the 
periastron.
}
\label{fig_sch}
\end{figure}

\begin{figure*}[t!]
\centering
\includegraphics[width=0.9\textwidth]{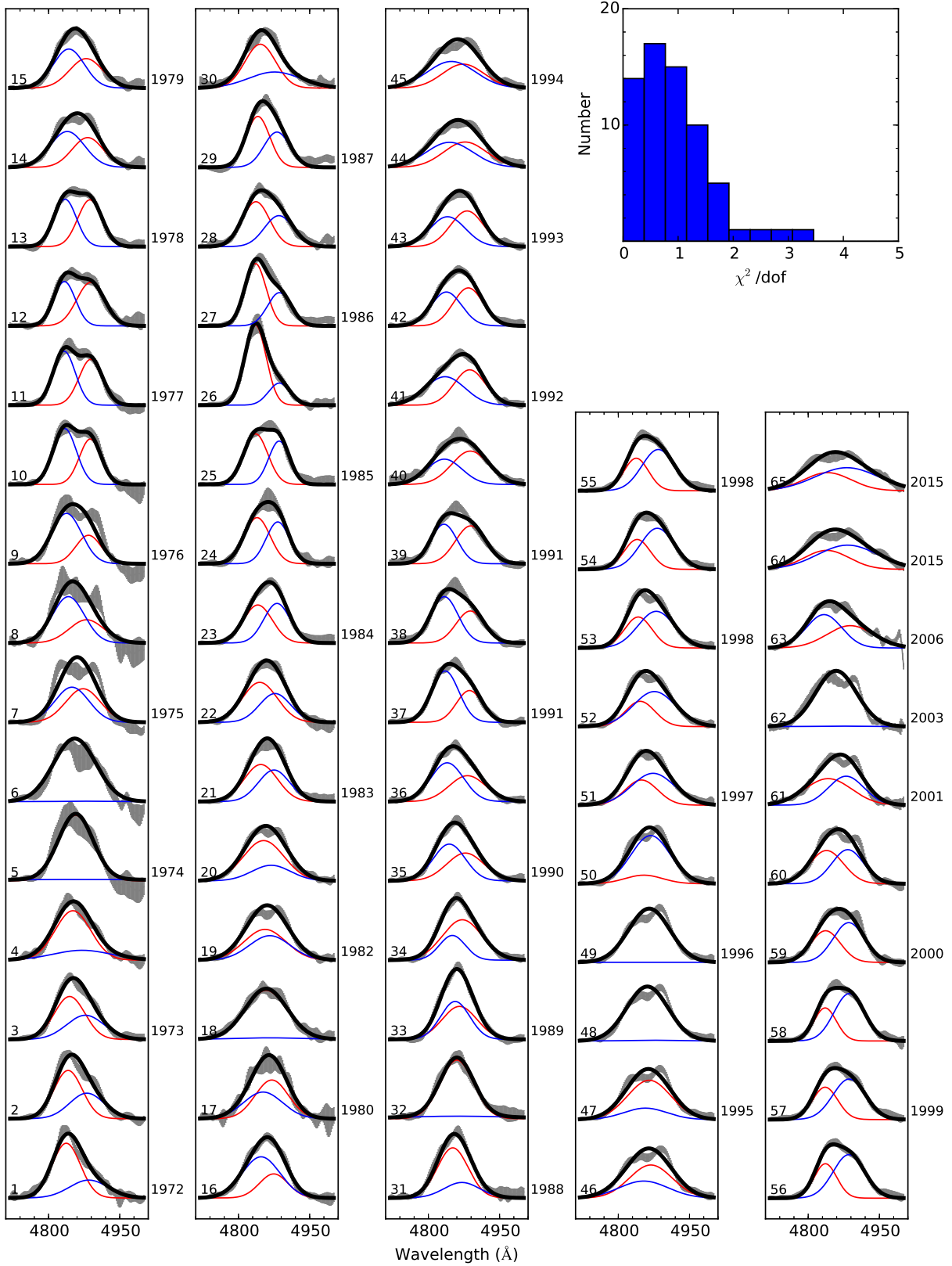}
\caption{\footnotesize
Best fit of the H$\beta$ spectral profiles, which are averaged over an interval 
of $150$ days. Blue and red lines are components from the binary BLRs, and 
black line is the sum of these two components. The top right inset shows the $\chi^2$ 
distribution of the best fit.}
\label{fig_fits}
\end{figure*}

\begin{deluxetable*}{clcc}
 \tablecolumns{4}
 \tablewidth{0.8\textwidth}
 \tablecaption{ Parameters of the supermassive black hole binary in NGC~5548}
 \tabletypesize{\footnotesize}
 \tablehead{
 \colhead{Parameter}       &
 \colhead{Physical Meaning}&
 \colhead{Value}           &
 \colhead{Unit}            
 }
\startdata
$\bhT$                  &Orbital period in rest frame                                      & $13.90\pm 0.85$        & yr          \\
$\bhm$                  &Total mass                                                        & $27.66\pm 8.66$        & $10^7\sunm$ \\
$\bha$                  &Semi-major axis                                                   & $21.73\pm 2.43$        & light-day   \\\hline 
$M_{\bullet,1}\sin^3i$  &Primary mass                                                      & $0.92^{+0.19}_{-0.24}$ & $10^7\sunm$ \\
$q$                     &Mass ratio                                                        & $1.00^{+0.00}_{-0.32}$ & \nodata     \\ 
$\bhe$                  &Eccentricity of the orbit                                         & $0.13^{+0.18}_{-0.13}$ & \nodata     \\
$\phi_0$                &Fraction of the orbit that the periastron occurs, prior to $t=0$  & $0.78^{+0.05}_{-0.06}$ & \nodata     \\ 
$\omega$                &Angle of the periastron with respect to X-axis                    & $202^{+20}_{-17}$      & deg         \\
$i$                     &Inclination of the orbit                                          & $23.9\pm 3.5$          & deg
\enddata
\tablecomments{ The total mass of the binary is determined from the $\bhm-\sigma_\star$ relation of
                \cite{Kormendy2013} for classical bulges. The orbital period is set by the rest-frame 
                mean period in Table~\ref{tab_period}.}
\label{tab_par}
\end{deluxetable*}

\subsection{SMBHB model}
Periodicity is one generic feature of SMBHB systems, not only in the orbital 
motion, but also in various observational properties related to the orbital motion.
Numerical simulations have extensively shown that the SMBHB's orbital motion
gives rise to enhanced periodicity in the mass accretion rates onto each black holes
(e.g., \citealt{Hayasaki2008, Hayasaki2013, Shi2012, Farris2014} and references
therein). This motivates systematic searches of SMBHB candidates through periodic
optical variability (e.g., \citealt{Graham2015b, Liu2015}).
On the other hand, the motion of the BLR gas surrounding the black holes 
is influenced, if not governed, by the binary, and the broad emission lines should
carry the information of the binary's orbital motion in their profiles 
(\citealt{Eracleous2012, Shen2010, Popovic2012, LiuJ2015}). Therefore, 
a SMBHB can naturally, simultaneously explain all the periodicity in NGC~5548, namely,
the periodicity in the 5100~{\AA} continuum, H$\beta$ integrated fluxes, and H$\beta$ profiles. 
Below we will develop a simple toy SMBHB model for the H$\beta$ profiles of 
NGC~5548 and discuss the implications for the connection between the continuum
variation and the orbital motion.

\section{A Toy SMBHB Model for the H$\beta$ Profiles}
With the above evidence that NGC~5548 has a SMBHB in its center, we now turn 
to explore the orbital information of the system. Below, we construct a simple toy
model for a binary black hole pair, each having its own accretion disk and 
BLR, and use this model to decompose the observed profiles of the H$\beta$ 
line (\citealt{Popovic2000, Popovic2012}). 

\subsection{Orbital parameters}
As the masses of the accretion disks and BLRs are negligible compared 
to those of the black holes, the orbital motion of the binary is simplified 
to a two-body problem.  We create a Cartesian coordinate system ($X, Y, Z$) 
with its origin at the center of mass of the binary. The orbital plane is 
located in the $XOY-$plane. Figure~\ref{fig_sch} illustrates a schematic 
for the geometry of the SMBHB system. In Appendix~D, we give a detailed 
derivation of equations for the orbital motion of the binary system. 
The binary orbit is completely described by seven parameters: 
\begin{itemize}
 \item mass of the primary black hole $M_{\bullet, 1}$,
 \item mass ratio $q$, 
 \item orbital eccentricity $\bhe$,
 \item orbital period $\bhT$ (rest-frame),
 \item angle $\omega$ of the periastron with respect to the $X-$axis,
 \item time fraction $\phi_0$ of the orbit (in term of the period) that the periastron 
 occurs, prior to the start of the observations ($t=0$),
 \item inclination angle $i$.
\end{itemize}
The inclination angle is degenerate with $M_{\bullet, 1}$ along lines 
of constant $M_{\bullet, 1}\sin^3i$. We use $M_{\bullet, 1}\sin^3i$ as a single 
parameter in the calculations and determine their respective values using the 
otherwise measured total mass of the binary (e.g., from the $\bhm-\sigma_\star$
relation). We also fix the orbital period by the {\it rest-frame} mean period 
listed in Table~\ref{tab_period}, i.e., $\bhT=14.14/(1+z)=13.90$~yr.

\begin{figure*}[t!]
\centering
\includegraphics[width=0.8\textwidth]{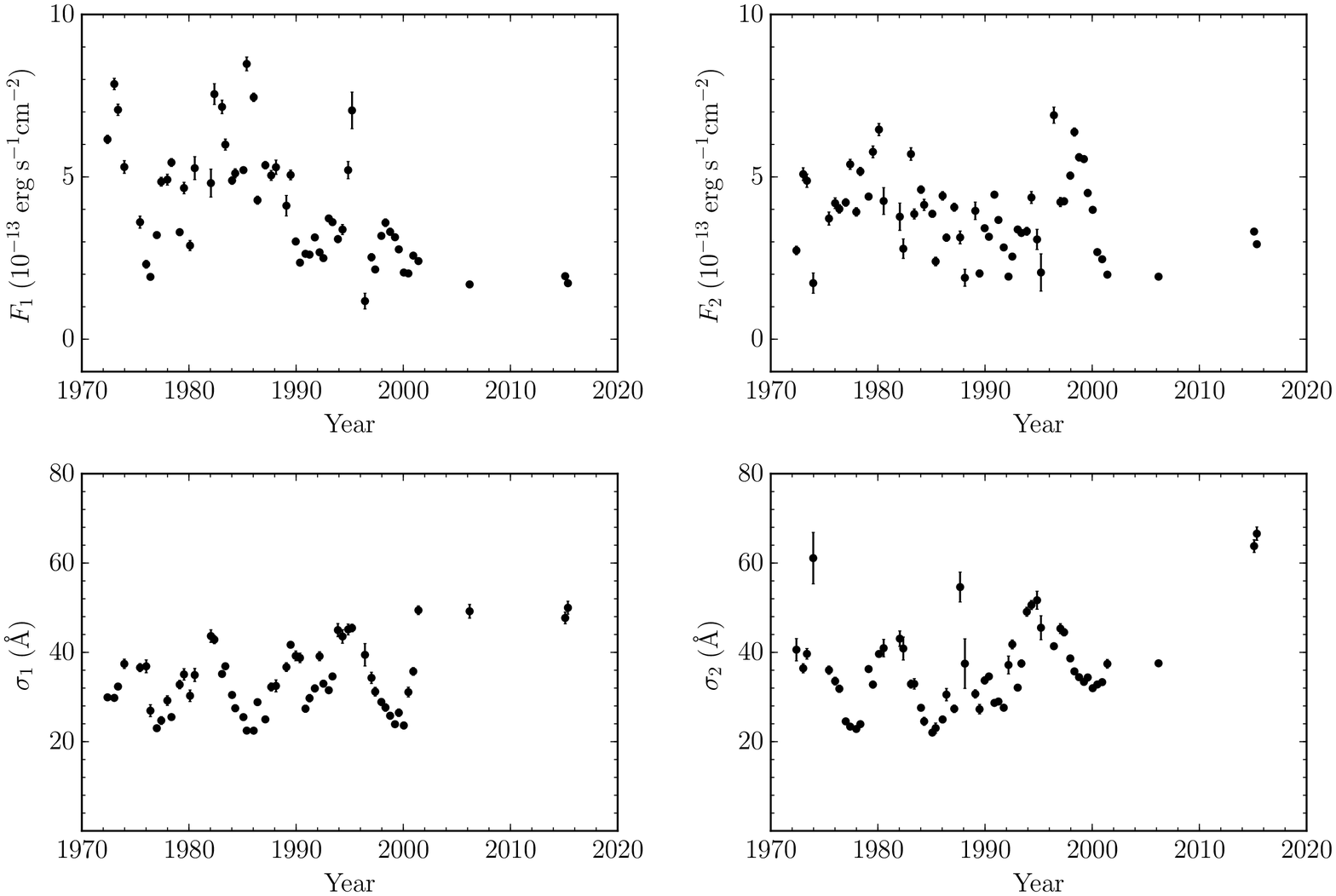}
\caption{\footnotesize
The best parameters $F_j$ and $\sigma_j$ for 65 averaged H$\beta$ profiles over time (see 
Equation (\ref{eqn_gau})). 
}
\label{fig_pars}
\end{figure*}

\subsection{Decomposition of H$\beta$ profiles}

To smear out the stochastic, short-term changes of the profiles, we average the 
H$\beta$ spectra over an interval of 150 days, chosen to be similar to the characteristic 
timescale in the light curves of continuum and H$\beta$ fluxes (see Section 3.1). 
Provided that the interval is much less than the orbital period, such an averaging 
procedure does not influence our results. As explained in Section~2.3, we 
pre-select out the spectra during relatively high states of NGC~5548 
($F_{\rm AGN}/F_{\rm gal}\gtrsim0.7$). This leaves us with 65 averaged H$\beta$ 
profiles, as plotted in Figure~\ref{fig_fits}.

We assume that the velocity profiles of the binary BLR components are Gaussian,
with shifting velocities following the line-of-sight velocities of their 
associated black holes, $V_{\rm LOS, 1}$ and $V_{\rm LOS, 2}$. To be specific, 
the profiles are described by
\begin{equation}
f_{j}(v)=\frac{F_j}{\sqrt{2\pi}\sigma_j}\exp\left[-
                \frac{(v-V_{{\rm LOS}, j})^2}{2(\sigma_j)^2}\right],
\label{eqn_gau}
\end{equation}
where $v$ is velocity, $j=1, 2$ correspond to the two components, and
$\sigma_j$ and $F_j$ are the velocity dispersion and total flux of the 
profile, respectively, which are the free parameters to be determined.
Note that $F_j$ and $\sigma_j$ are different with different H$\beta$ spectra
(since the H$\beta$ line is time-varying), so that we have hundreds of 
free parameters, making a simultaneous determination of them impossible.
Our decomposition procedure is as follows. Given a set of orbital 
parameters, we first calculate the line-of-sight velocities of the binary 
and then use Equation~(\ref{eqn_gau}) to fit each averaged H$\beta$ spectrum
based on $\chi^2$ minimization. The minimum $\chi^2$ of each fit is denoted 
by $\chi_{i, \rm min}^2$; there are in total 65 $\chi_{\rm min}^2$. 
We sum up these $\chi_{i,\rm min}^2$ as 
\begin{equation}
\chi_{\rm tot}^2 = \sum_i \chi_{i,\rm min}^2
\end{equation}
and determine the best orbital parameters by searching for the 
minimum $\chi_{\rm tot}^2$. As an independent check, we expect that 
the obtained parameters $F_j$ and $\sigma_j$ should also change periodically 
with time.

It is computationally expensive to fit 65 averaged H$\beta$ profiles 
every time given a new set of orbital parameters, meaning that 
we cannot efficiently explore all the parameter space to find out 
the minimum $\chi_{\rm tot}^2$. Instead, we only survey over a mesh 
grid of the probable orbital parameters. We can impose prior constraints on
the parameters from general considerations. The range of mass ratio is set 
to $q\sim[0.3, 1.0]$, as the morphology of NGC~5548 suggests it experienced
a major merger event (Section 1). The orbit eccentricity must be small, or else
the binary will spend most its time around the apastron. In this case, 
most of the H$\beta$ profiles will not be as asymmetric as observed 
(see Figure~\ref{fig_phase}). We accordingly set a range of eccentricity 
$\bhe\sim[0,0.5]$. The inclination angle should be not larger than the opening 
angle of the dusty torus, or else the BLRs will be obscured (e.g., \citealt{Lawrence2010}). 
We set the range of inclination angle $i\lesssim50^\circ$. This yields an 
upper limit to $M_{\bullet,1}\sin^3i\lesssim9\times10^7M_\odot$. The ranges 
for the rest of the free parameters are $\omega\sim[0, 2\pi]$ and $\phi_0\sim[0, 1]$. 
After determining the best-fit orbital parameters, we estimate 
the associated uncertainties from the 68.3\% confidence ranges (``$1\sigma$'') 
of the $\chi^2_{\rm tot}$ curves. The resulting best-fit orbital parameters are 
tabulated in Table~\ref{tab_par}. The binary has a large mass ratio close 
to 1 ($q>0.68$), consistent with the major merger scenario for the tidal 
morphology of the host galaxy. The orbit has an eccentricity of $\bhe<0.31$ 
and an inclination angle of $23.9\pm3.5$ degrees.

In Figure~\ref{fig_fits}, we show the best fits of the averaged profiles 
and the corresponding $\chi^2$-distribution. Generally, considering the 
simplicity of our decomposition model, most of the H$\beta$ profiles are 
well reproduced, except for a few with tiny lumps during some epochs 
(e.g., 1974-1977 and 2000-2003). These tiny lumps may arise from over-dense 
regions in the BLRs due to the tidal perturbations of the binary, which may 
induce inhomogeneities in the BLR gas distribution (e.g., \citealt{Bogdanovic2008}).
Moreover, when the line-of-sight velocities of the two components are 
nearly identical (both go to zero), the fitting is degenerate and 
only finds a solution for one component. This is why in panels 5, 6, 18, 32, 48, 49, 
and 62 the obtained flux of one component is zero.
Figure~\ref{fig_pars} plots the best-fit parameters $F_j$ and $\sigma_j$ for 
the 65 H$\beta$ profiles, excluding the points whose fits are degenerate 
due to nearly identical line-of-sight velocities of the two components. 
As can be seen, it is no surprise that there appears 
to be some periodicity in these parameters over time, which just reflects that 
the H$\beta$ profiles have periodic variations, as we have demonstrated 
through the epoch-folding scheme.
It is worth mentioning that the variability of the line dispersion 
of each component is due to the changes in the ionizing luminosity of the disk(s), 
as reflected in the 5100~{\AA} light curve.

\begin{figure}[t!]
\centering
\includegraphics[angle=0,width=0.45\textwidth]{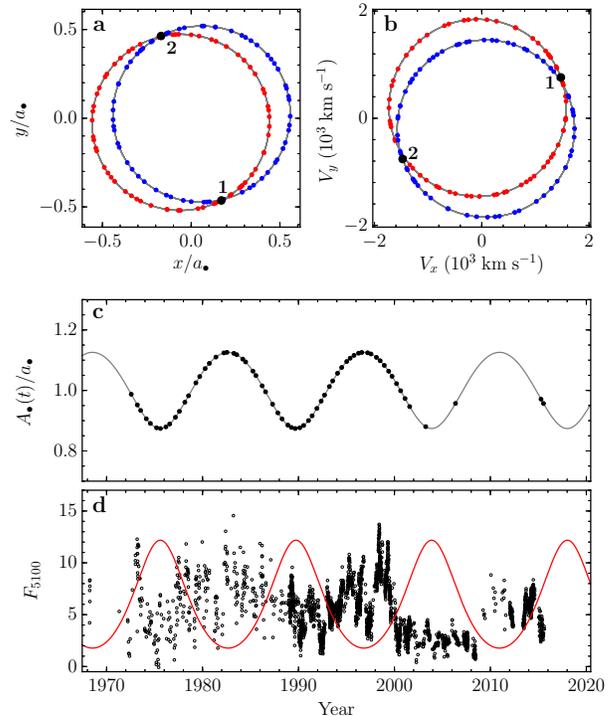}
\caption{\footnotesize The ({\it a}) orbit, ({\it b}) velocity, ({\it c}) 
separation between the black hole binary in the orbital plane, and ({\it d}) 
relation between the continuum fluxes and binary separation. Small dots in panels 
{\it a-c} represent the points associated with the averaged spectra as shown 
in Figure~\ref{fig_fits}. Large black dots labelled ``1" and ``2" in panels 
{\it a} and {\it b} mark the initial positions and velocities of the binary. 
In panel {\it d}, for illustrative purposes, we superpose a red curve 
$F_{5100}=a(A_\bullet/a_\bullet)^{-2}+b$, with $a=20$ and $b=-14$,
to compare the 5100~\AA\, light curve with the orbital motion. There appears to be a 
time offset of a few years between the orbital motion (pericenter passage) and the 
continuum variations.} 
\label{fig_orbit}
\end{figure}

\subsection{The continuum variations lag behind the orbital motion?}
Figures \ref{fig_orbit}{\it a}$-$\ref{fig_orbit}{\it c} 
show the trajectories, velocities, and separation of the SMBHB system in the 
orbital plane deduced from our model.  To compare the 5100~{\AA} light curve 
with the orbital motion, in Figure~\ref{fig_orbit}{\it d} we superpose a 
red curve to indicate the inverse square of the binary's separation 
as $a(A_\bullet/a_\bullet)^{-2} + b$, with $a = 20$ and $b = -14$.
The choice of these values is arbitrary and just for illustrative purposes.
There appears to be a time offset of a few years between the orbital 
motion (pericenter passage) and the continuum variations. 

We suspect that such a time offset may be due to the (different) retarded responses of 
the accretion disks to their interplay with the binary. The interplay 
between the disks and the binary includes: (1) perturbations in the disks 
induced by the tidal torques; (2) modulation of the mass accretion rates 
onto the disks by the orbital motion of the binary. 
The latter is well studied by numerical simulations (e.g., \citealt{Farris2014}, 
and references therein). At the apastron (rather than periastron) passage, 
the black holes are closest to the inner edge of the circumbinary disk
and therefore are easiest to obtain the gas%
\footnote{The mass accretion rate from the inner edge of the circumbinary disk 
onto the two black holes is not necessarily equal to the rate in the circumbinary disk, 
since the tidal torque raised by the binary may lead to gas pileup around the inner 
edge of the circumbinary disk (e.g., \citealt{Rafikov2013}
and references therein).}.
If further taking into account the free-fall time for the gas streaming 
from the inner edge of the circumbinary disk to the black holes, there will 
be a time delay between the peak of the mass accretion rates onto the black holes
and the apastron passage (e.g., \citealt{Artymowicz1996, Hayasaki2013}). Moreover,
the system is now no longer stationary but highly time dependent;
but the viscous accretion timescale, which reflects the timescale 
of an accretion disk responding to changes in mass accretion rate, is 
on the order of $10^{5-7}$ yr (\citealt{Frank2002}), much longer that 
the modulation timescale of mass accretion. Therefore, it is inappropriate 
to directly use mass accretion rate to indicate disk emission.
The dynamics of accretion disks with black hole tidal interaction may have
special observational properties, which deserves a dedicated investigation in 
the future (see the preliminary studies of \citealt{Tanaka2013} and
\citealt{Farris2015}). 

\section{Discussions on the SMBHB scenario}
The intensive spectroscopic and photometric data of NGC~5548 over a time span of four decades 
allow us to tightly constrain the possible explanations for the periodicity 
found in the light curves of the 5100~{\AA} continuum and H$\beta$ flux, as well as in the variations of the 
H$\beta$ profile. The SMBHB scenario can naturally account for the three types of periodic 
variations observed in NGC 5548. However, the following points merit further thought.

Although we ascribe the periodicity in the light curves of the continuum
and H$\beta$ fluxes to binary motion, we cannot specify in more detail 
how the (periodic) binary motion translates into periodic luminosity 
modulation. A general, qualitative explanation is that the  binary's orbital
motion induces periodic enhancements of the mass accretion rate onto each 
black hole (e.g., \citealt{Farris2014}). As illustrated in Section~5.3, it is 
not as straightforward as usually thought to link the disk emission with mass 
accretion rate. One needs to carefully investigate the dynamics of accretion 
disks around binary black holes and the influences of the tidal torques 
(e.g., \citealt{Liu2010}). On the other hand, recently \cite{Dorazio2015} 
interpreted the periodicity of the light curve by relativistic Doppler 
boosting and successfully applied their model to the periodic light curve of PG~1302-102 
(\citealt{Graham2015}). For NGC~5548, Doppler boosting is unimportant because 
the line-of-sight velocities of the binary are non-relativistic ($V/c<10^{-2}$; 
see Figure~\ref{fig_orbit}b).

We directly interpret the observed periodicity as the (redshifted) orbital 
period of the binary. However, recent hydrodynamic numerical simulations 
tracked mass accretion onto the binary and showed that there may be multiple periodic 
components of variability in accretion rate (e.g., \citealt{Farris2014, 
Charisi2015, Dorazio2015a}), which may lead to periodic variations in continuum flux. 
In some cases, depending on the mass ratio of the binary, the dominant 
periodicity may not correspond to the orbital period.
However, we do not yet know whether the variations of the H$\beta$ 
profiles have similarly multiple periodic components. 

For simplicity, we use two Gaussians to delineate the H$\beta$ emission 
lines from the BLR around each black hole, assuming that the two BLRs are well separated and virialized.
On the one hand, tidal torques from the 
binary induce perturbations to the BLR gas, giving rise to some inhomogeneous 
features, such as tidal arms or clumpy filaments (\citealt{Bogdanovic2008}). 
Clearly, these features, which may be responsible for tiny lumps appearing in 
the H$\beta$ profiles (see Section~5.2), are time varying and closely linked to 
the binary orbit. On the other hand, the semi-major axis of the binary is 
$\sim22$ light-days, only marginally larger than or even comparable with the BLR sizes 
measured from the reverberation-mapping observations (\citealt{Bentz2013}). 
The Roche lobes of the binary system have a size of
(\citealt{Eggleton1983})
\begin{equation}
\frac{R_{\rm L}}{a_\bullet} = \frac{0.49q^{2/3}}{0.6q^{2/3}+\ln(1+q^{1/3})}\sim 0.4,
\end{equation}
for a circular orbit ($\bhe=0$) with equal mass ($q=1$). 
The size of the BLR associated with each black hole can be estimated from 
the $R_{\rm BLR}-L$ relation of \cite{Bentz2013} as
\begin{equation}
 R_{\rm BLR} = 33.6\left(\frac{L_{\rm 5100}}{10^{44}~{\rm erg~s^{-1}}}\right)^{0.533}
 \sim 10~\text{light-days},
\end{equation}
where we assume that the black holes in the pair have the same luminosity and 
adopt the typical mean value $L_{\rm 5100}=10^{43}~{\rm erg~s^{-1}}$ using
a luminosity distance of 75~Mpc.
Considering that the 5100~{\AA} luminosity varies by a large factor of more than 6 (see Figure~\ref{fig_lc}),
the above estimates indicate that the two BLRs are plausibly in contact
during the intermediate or high states of NGC~5548 (in which $R_{\rm BLR}>R_{\rm L}$).
In such cases, there probably will be a circumbinary BLR component. 
Meanwhile, a portion of the two BLRs may be illuminated by ionizing photons
from both black holes (\citealt{Shen2010}). Therefore, a Gaussian is only a very
simplified approximation to the realistic H$\beta$ profiles
and {\it the obtained orbital parameters (such as the mass ratio, 
eccentricity, and inclination, etc.) should be taken with caution.} 
These values are meant to demonstrate the capability of reproducing the 
H$\beta$ profiles with a SMBHB model. More 
sophisticated SMBHB models are required to account for the above mentioned effects.

Finally, we present an outlook for subsequent works and future observations 
to further reinforce the presence of a SMBHB system in NGC~5548, with the following remarks. 

\begin{enumerate}

\item If each black hole produces radio emission, the future Event Horizon Telescope, 
with an angular resolution of $\sim 20\,\mu$as (\citealt{Fish2013}), will be capable 
of spatially resolving the radio core of NGC~5548 (\citealt{HU2001}) into two components 
separated by $\sim 22$ light-days ($\sim 50\,\mu$as).  This will provide direct evidence for the SMBHB. 

\item We only apply our SMBHB model to decompose the averaged H$\beta$ profiles so as to smear 
out the short-term stochastic changes in the profiles. It would be interesting to study 
in detail how the H$\beta$ profile varies on short timescale of hundreds of days 
(e.g., \citealt{Flohic2008}). As demonstrated by \cite{Shen2010} using a heuristic toy model, 
the reverberation behavior of the broad emission lines from a binary system should be distinguishable 
from that from a single black hole, considering that the BLR geometry is different and there are two 
ionizing sources in binaries.  Such studies will also shed light on the BLR dynamics surrounding a SMBHB.

\item From the extensive database for NGC~5548, we can also use other prominent emission lines 
(such as H$\alpha$) to study their profile variations by applying the same procedure as in this work 
and test if there exists similar periodicity.  

\item We can use $N$-body simulations to recover the merger history of NGC 5548 by trying to 
reproduce in detail its observed morphology.  This will help to better understand the dynamical 
evolution of the SMBHB (e.g., \citealt{Begelman1980}). 

\item Continued monitoring of NGC~5548 will track more cycles of the binary orbit and improve 
the measurement uncertainties of the periods.  
\end{enumerate}

\section{Conclusions}
We propose that a SMBHB plausibly resides in the galactic center of NGC 5548 
based on four decades of spectroscopic data. The two peaks of the H$\beta$ profiles 
shift and merge in a systematic manner with a period of $\sim 14$ yr, which 
agrees with the period observed in the long-term variations in both the continuum and H$\beta$ 
fluxes (see Figures \ref{fig_lc} and \ref{fig_phase}). In addition, the 
morphology of NGC~5548 shows two long tidal tails (see Figure~\ref{fig_img} 
and \citealt{Tyson1998}), indicative of a major merger event that occurred $\sim 1$\,Gyr ago. 
These lines of observations make NGC~5548 one of the nearest and best sub-parsec 
SMBHB candidates known to date.
The SMBHB has a total mass of $2.8\times10^8M_\odot$ from the $\bhm-\sigma_\star$ 
relation and a semi-major axis of 21.73 light-days from Kepler's third law, 
indicating that the SMBHB has an extremely sub-parsec separation. 
Using our toy SMBHB model, we demonstrate that the complex, secularly varying 
H$\beta$ profiles  can be reproduced by orbital motion of the SMBHB with 
a large mass ratio ($q>0.68$) and an eccentricity of $\bhe<0.31$, viewed 
at an inclination angle of 23.9 degrees (see Table \ref{tab_par}).

The SMBHB will coalesce within $\sim 6\times 10^6a_{-2}^4m_8^{-3}q^{-1}(1+q)^2$ yr 
and radiate gravitational waves at a frequency of
$2/T_\bullet\sim 7\times 10^{-9}m_8^{1/2}a_{-2}^{-3/2}$~Hz, where $a_{-2}=\bha/10^{-2}$ pc. 
The expected strain amplitude of the intrinsic gravitation wave is 
$h_s\approx 9\times10^{-17}m_{8}^{2}a_{-2}^{-1}q(1+q)^{-2}d_{75}^{-1}$
(e.g., see \citealt{Graham2015b}), where $d_{75}=d_{\rm L}/75\,{\rm Mpc}$ is the 
distance to the observer. The proximity of NGC~5548 makes its SMBHB system an 
excellent target for detection of gravitational waves through the Pulsar Timing Array
(e.g., \citealt{Moore2015, Sesana2015}).

\acknowledgements
We thank the referee for constructive suggestions that significantly improved the 
manuscript and Yue Shen for useful discussions and pointing out that spiral arms 
may trigger periodic accretion. We are grateful to the staff of 
the Lijiang station of the Yunnan Observatories
for their assistance with observations, and to S. G. Sergeev, T. Iijima, K. Denney, 
and L.~{\v C}. Popovi\'c for their kind help in providing data.
This research is supported in part by the Strategic Priority Research Program - 
The Emergence of Cosmological Structures of the Chinese Academy of Sciences, 
Grant No. XDB09000000, by NSFC grants NSFC-11173023 (C.H.), -11133006 (J.M.B.), 
-11233003 (J.M.W.), -11303026 (P.D.), -11573026 (Y.R.L.), -U1431228 (J.M.W), and -11361140347 (J.M.B.),
by the Key Research Program of the Chinese Academy of Sciences, Grant No. KJZD-EW-M06 (J.M.B.), 
and by the Kavli Foundation and Peking University (L.C.H.). This work has made use of data from 
the Lick AGN Monitoring Project public data release 
(\url{http://www.physics.uci.edu/~barth/lamp.html}).

\bibliographystyle{plain}


\appendix

In this Appendix, we present detailed information on the published data 
used in our analysis, a study on the host galaxy contamination on H$\beta$ profiles, a test 
of the influences of the varying H$\beta$ line width (due to the periodic variations in the 
5100~{\AA} continuum) on the epoch-folding results, and 
a derivation of the SMBHB orbital motion used in our model for the H$\beta$ profiles.

\section{A. Historical observations}

\begin{figure*}[t!]
\centering
\includegraphics[angle=270,width=0.8\textwidth]{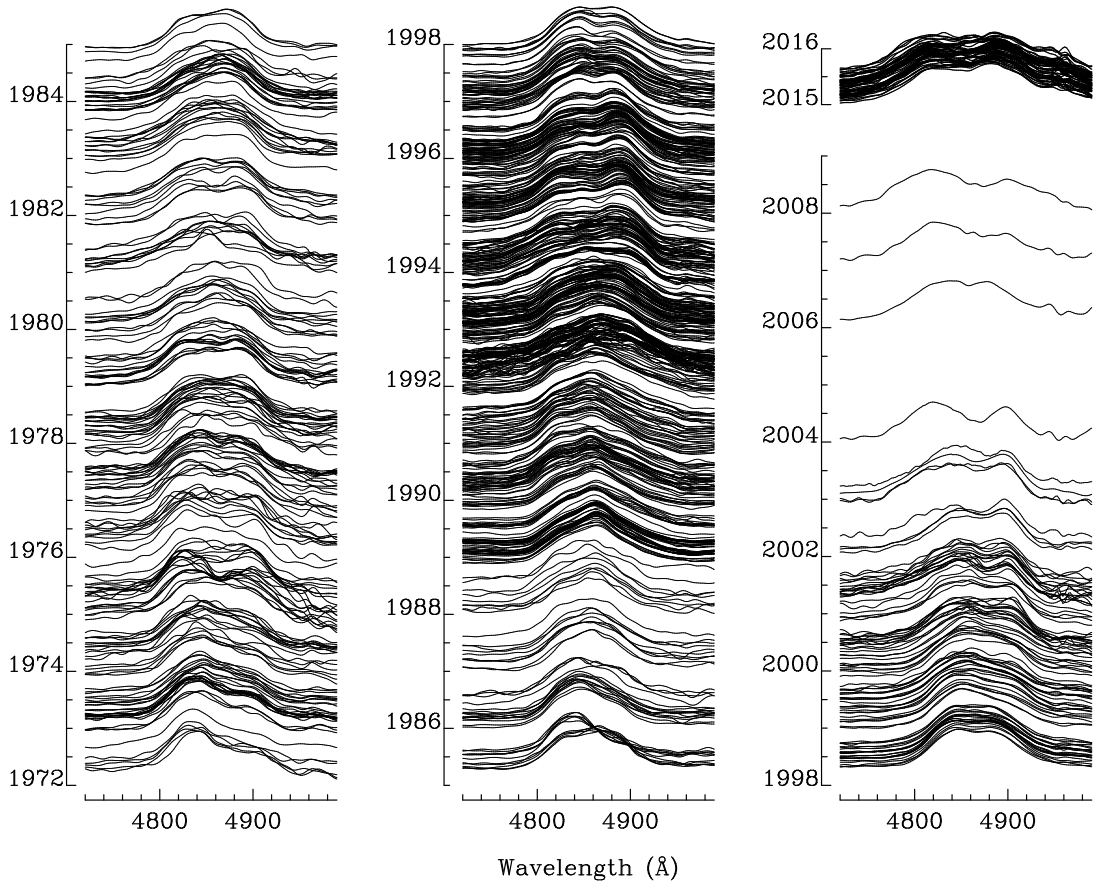}
\caption{\footnotesize A total of 924 H$\beta$ profiles of NGC 5548 from 1972 to 2015. } 
\label{fig_spec}
\end{figure*}

\subsection{A.1. Data sources}
We compiled the spectroscopic and photometric data for NGC~5548 from the literature 
in Tables \ref{tab_hb} and 
\ref{tab_con}.  In addition, as described Section 2.1, we made new 
observations of our own in 2015.  The majority of the data are publicly 
available. The others are kindly provided by the corresponding authors of the 
listed references.  The archival data of the 5100~\AA\ continuum fluxes and 
the H$\beta$ profiles from 1972 to 2001 compiled in \cite{Sergeev2007} were 
provided by S. G. Sergeev.  The spectra from 1995 presented in 
\cite{Iijima1995} were provided by T. Iijima, and those from 1998--2004 
presented in \cite{Popovic2008} were provided by L.~{\v C}. Popovi{\'c}. The mean spectra 
from 2008 presented in \cite{Denney2010} was provided by K. 
Denney. Figure~\ref{fig_spec} shows a total of 924 H$\beta$ profiles taken over four decades.

\subsection{A.2. Flux conversion}

Some studies only presented continuum measurements in the $V$ band or at 
1350 \AA.  To convert these data into continuum flux densities at 5100~{\AA}, 
we use the following relations:
\begin{itemize}
\item The $V$-band flux density, $F_{\lambda}(V)$, can be converted to 
5100~\AA\ flux density through
\begin{equation}
F_{5100}=(0.946\pm 0.072)F_\lambda(V),
\end{equation}
as determined from simultaneous observations (\citealt{Romanishin1995}).

\item For photometric observations in the $V$ band, we first convert the 
magnitudes into flux densities by adopting the zero point $F_\lambda(V=0)=
3.92\times10^{-9}\,{\rm erg~s^{-1}~cm^{-2}}\textrm{\AA}^{-1}$ 
(\citealt{Johnson1966}) and then apply relation~(A1).

\item The flux density at 1350 {\AA} can be converted to 5100~\AA\ through 
the relation 
\begin{equation}
\log\left(\frac{L_{5100}}{10^{43}~{\rm erg~s^{-1}}}\right) =(0.55\pm0.05) +
(0.63\pm0.12)\log\left(\frac{L_{1350}}{10^{44}~{\rm erg~s^{-1}}}\right),
\end{equation}
which was derived from simultaneous observations in the UV and optical bands
(\citealt{KilerciEser2015}).  Note that this relation is based on a luminosity 
distance of $72.5\pm7.0$ Mpc for NGC~5548.
\end{itemize}

\begin{figure*}[t!]
\centering
\includegraphics[width=0.8\textwidth]{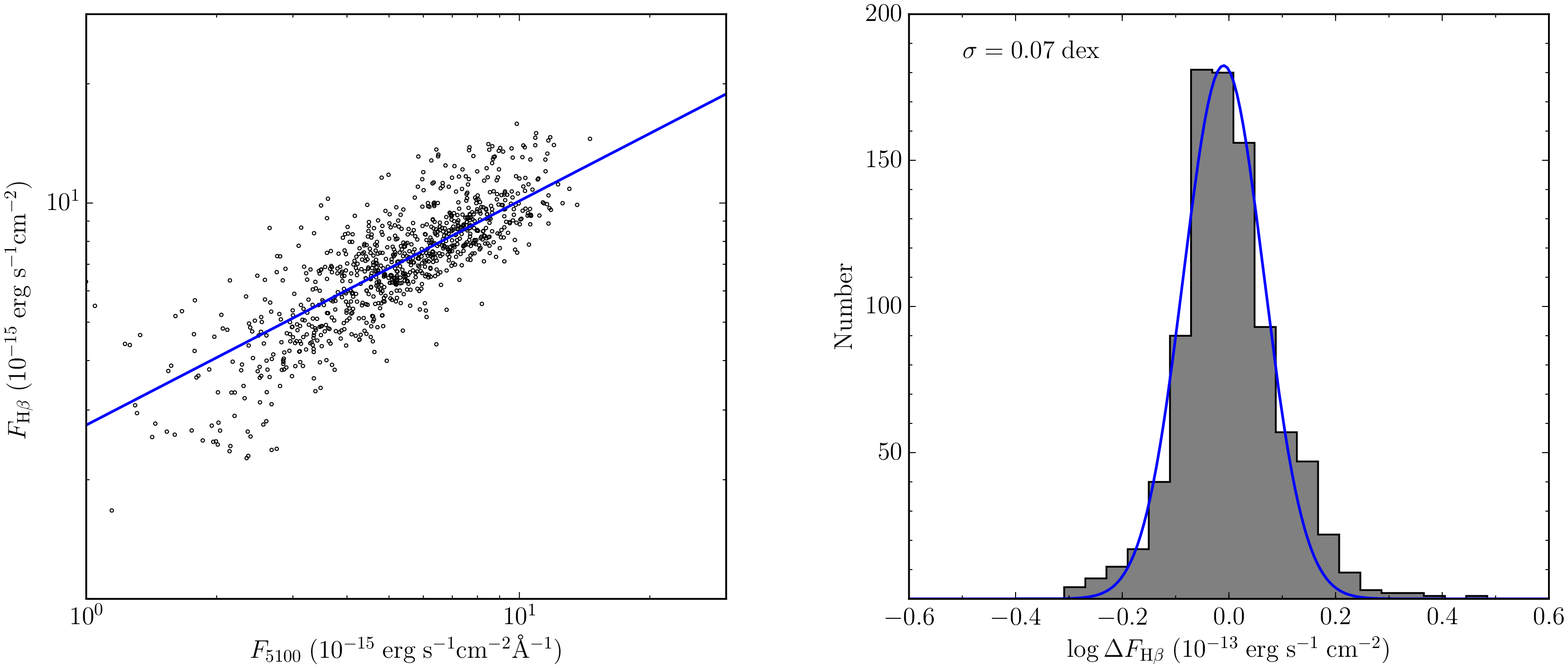}
\includegraphics[width=0.8\textwidth]{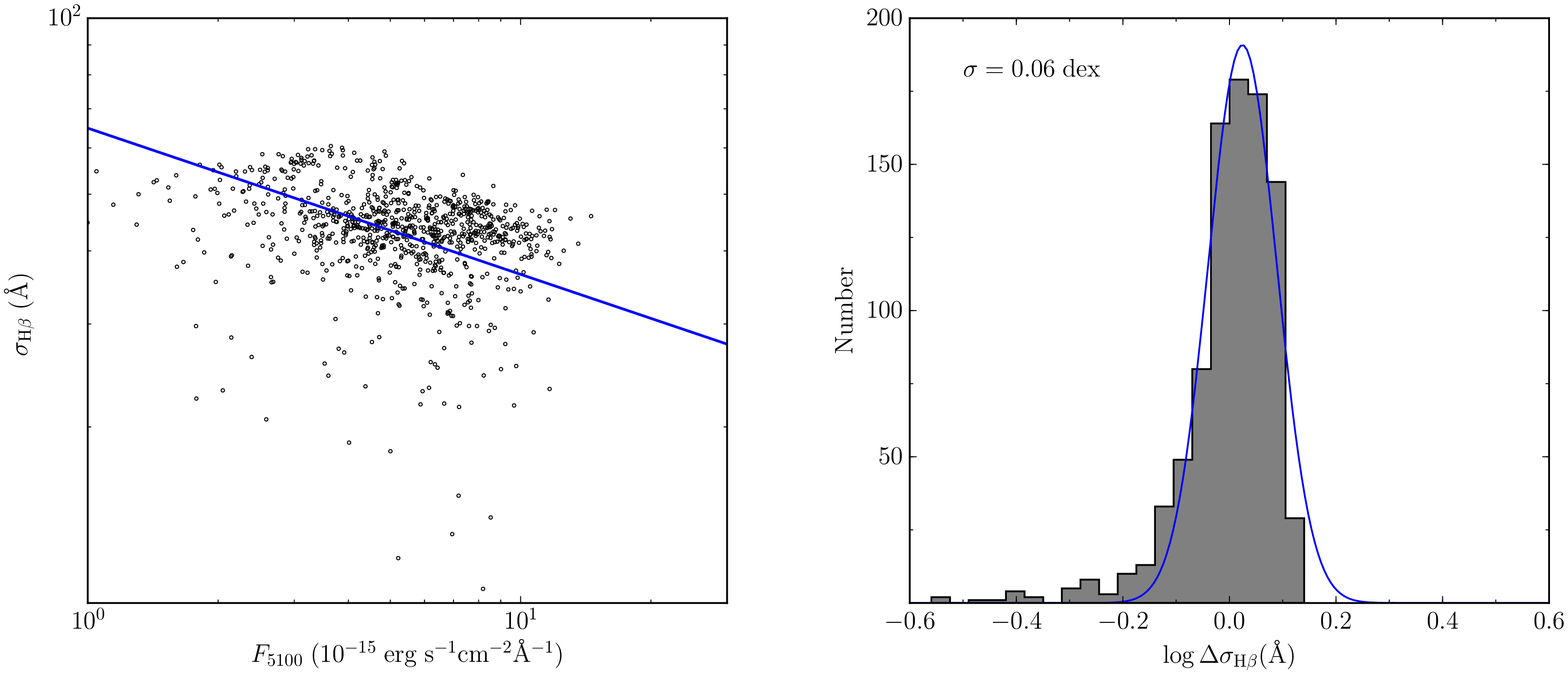}
\caption{Left: relations between the 5100~{\AA} continuum flux and (top) the H$\beta$ line flux and 
(bottom) the H$\beta$ line dispersion. The narrow H$\beta$ component is subtracted and the H$\beta$
flux is calculated over a wavelength window 4719-5001~{\AA} in the rest frame. Solid line
represents the best fit to the data. Right: residuals of (top) the H$\beta$ line flux and 
(bottom) the H$\beta$ line dispersion using the corresponding best-fit relation. 
Solid line represents a fit to the residuals by a Gaussian function.}
\label{fig_relation}
\end{figure*}

\section{B. Host contamination of the H$\beta$ profile}
In this section, we test whether contamination from host galaxy starlight 
causes the double-peaked profile of H$\beta$ that appears in most of the 
observations.  Following \cite{Barth2013}, we adopt an 11 Gyr, solar 
metallicity, single-burst spectrum from \cite{Bruzual2003} to model the host 
galaxy starlight. This template is convolved with a Gaussian with a standard deviation of
$\sim15$ {\AA} to account for the spectral resolution of the data. 
We assume that the optical continuum of the AGN can be described by a power law, and 
that the broad H$\beta$ profile is a single Gaussian.  The H$\beta$ flux is related to 
the continuum according to the following relation, which is empirically determined from 
the data compiled in this work:
\begin{equation}
\log\left(\frac{F_{{\rm H}\beta}}{10^{-13}\,\rm erg~s^{-1}~cm^{-2}}\right) = 
0.44 + 0.57\log\left(\frac{F_{5100}}{10^{-15}\,\rm erg~s^{-1}~cm^{-2}~{\text{\AA}}^{-1}}\right).
\end{equation}
The obtained coefficients are generally consistent with \cite{Sergeev2007}'s results.
The dispersion of the Gaussian for H$\beta$ depends on the continuum as
\begin{equation}
\log\left(\frac{\sigma_{_{\rm H\beta}}}{\text{\AA}}\right)
=1.81 - 0.25 \log\left(\frac{F_{5100}}{10^{-15}\,\rm 
erg~s^{-1}~cm^{-2}~{\text{\AA}}^{-1}}\right),
\label{eqn_sig}
\end{equation}
again based on the data used in this work. The scatter of the latter 
fit is about 0.06~dex, which we will use in next section. In Figure~\ref{fig_relation}, 
we show the relations between the 5100~{\AA} continuum flux and  the H$\beta$ line flux and 
the H$\beta$ line dispersion. For simplicity, the power-law index is 
fixed to $-$0.25, as expected from simple photoionization theory.  We generate 
a mock spectrum by adding up the three components, namely the host starlight, 
the AGN continuum, and the H$\beta$ line.  Since we only focus on the effect of 
host galaxy contamination on the broad H$\beta$ profile,  we neglect other components 
such as narrow H$\beta$, \oiii, Balmer continuum, and \feii\, blends. 

We set the host galaxy flux at 5100~{\AA} to the value determined from the 
AGN Watch campaigns, $F_{\rm gal}=3.75\times10^{-15}\,\rm erg~s^{-1}~cm^{-2}~
{\text{\AA}}^{-1}$ (\citealt{Peterson2013}).  Note that this choice does not affect 
our test because the results only depend on the flux ratio between the central AGN and 
the host galaxy starlight $F_{\rm AGN}/F_{\rm gal}$ at 5100~{\AA}.  We then use 
the standard procedure from \cite{Sergeev2007}, as described in Section 2.3, to remove 
the continuum of the mock spectrum and isolate the broad H$\beta$ line.  
The top panel of Figure \ref{fig_hbetahost} shows the resulting broad H$\beta$ profiles for 
different values of $F_{\rm AGN}/F_{\rm gal}$.  As can be seen, the broad 
H$\beta$ profile is significantly contaminated by host galaxy light when 
$F_{\rm AGN}/F_{\rm gal}\lesssim0.5$.  We also calculate the integrated H$\beta$ flux 
over a wavelength window $4714-4933$~{\AA} (rest-frame) and show its dependence on
$F_{\rm AGN}/F_{\rm gal}$ in the bottom panel of Figure~\ref{fig_hbetahost}.
The changes in the H$\beta$ fluxes are less than 10\% for $F_{\rm AGN}/F_{\rm gal}\gtrsim0.2$.

\begin{figure*}[t!]
\centering
\includegraphics[width=0.4\textwidth]{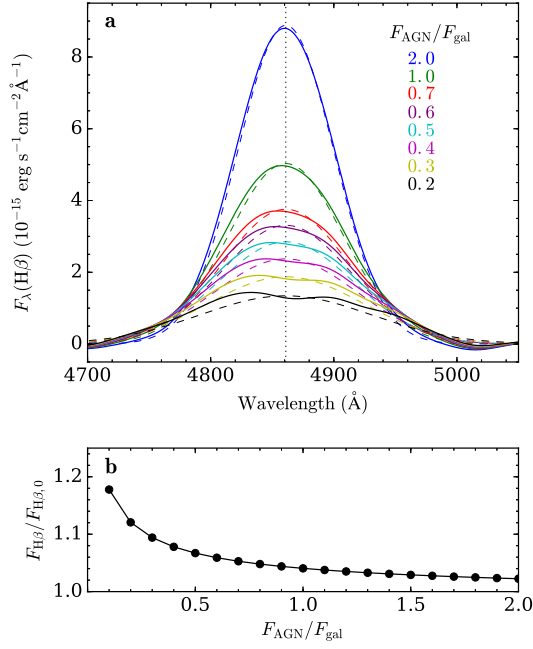}
\caption{\footnotesize
 The effect of host galaxy contamination on ({\it a}) the profile of H$\beta$ 
 and ({\it b}) H$\beta$ flux for different ratios between
 the AGN flux and the host galaxy starlight ($F_{\rm  AGN}/F_{\rm  gal}$) at 5100~{\AA}. 
 The host galaxy starlight is set to a typical value of
 $ F_{\rm gal}=3.75\times10^{-15}~\rm erg~s^{-1}~cm^{-2}~\text{\AA}^{-1}$.
 In the top panel, solid lines represent the obtained H$\beta$ profiles from the mock 
 spectra and dashed lines the input H$\beta$ profiles (see the text for details). 
 Vertical line shows the wavelength center of H$\beta$ line. In the bottom panel, 
 vertical axis is the ratio between the obtained ($F_{\rm H\beta}$) and input 
 integrated H$\beta$ fluxes ($F_{\rm H\beta, 0}$) .}
\label{fig_hbetahost}
\end{figure*}

\section{C. Influences of the Periodic 5100~{\AA} Variations on the 
Epoch-folding of the Broad H$\beta$ Profiles}
We note that simply from the relation between BLR sizes and AGN 5100~{\AA} 
luminosity and the virial relation (e.g., \citealt{Peterson2004, Bentz2013, Simic2016}), 
the H$\beta$ line width is closely correlated to the 5100~{\AA} luminosity.
As a result, the periodic continuum variations of NGC~5548 certainly
lead to periodic changes in H$\beta$ line width.
To see how the line width of H$\beta$ profile (in terms of the line dispersion) 
influences the epoch-folding 
results, we simulate mock H$\beta$ profiles using a single Gaussian, as follows.
\begin{itemize}
 \item Set the center of the Gaussian to be 4861 {\AA} (in the rest frame).
 \item Set the dispersion $\sigma_{\rm H\beta}$ using Equation 
(\ref{eqn_sig}), with a scatter of 0.06 dex.
\item Set the noise of the Gaussian at each wavelength bin by the standard 
deviation of the observation data over all the epochs at the same wavelength 
bin.
\end{itemize}
Figure~\ref{fig_sim}a shows the epoch-folding results 
for the mock data. Compared with the epoch-folding results for the real 
data in Figure~\ref{fig_phase},  the simulated  profiles 
show a similar periodicity of $\sim$14 yr, directly owing to the 
periodicity in the light curve of 5100~{\AA} continuum.

However, for the simulation data, we can just make a simple transformation 
for the H$\beta$ profiles to eliminate the influences of the varying line 
dispersion,
\begin{equation}
\frac{\lambda - \lambda_0}{\sigma_{\rm old}} \rightarrow \frac{\lambda - 
\lambda_0}{\sigma_{\rm new}},
\end{equation}
where $\lambda_0$ is the wavelength center of H$\beta$ profiles and 
$\sigma_{\rm old}$ and $\sigma_{\rm new}$ are the original and new line 
dispersion, respectively. We transform all the simulated H$\beta$ profiles to 
$\sigma_{\rm new}=50~\text{\AA}$. Similarly, we can also perform the same 
transformation for the observation data. 
Figures \ref{fig_sim}b and \ref{fig_sim}c show the comparison of the epoch-folding results by 
including the 
transformation. As expected, the periodicity disappears for the simulated data 
but is still present in the observation data, indicating that the observed periodicity is not 
entirely an artifact of the periodicity in the 5100~{\AA} continuum.

\begin{figure}[t!]
\centering
\includegraphics[width=0.3\textwidth]{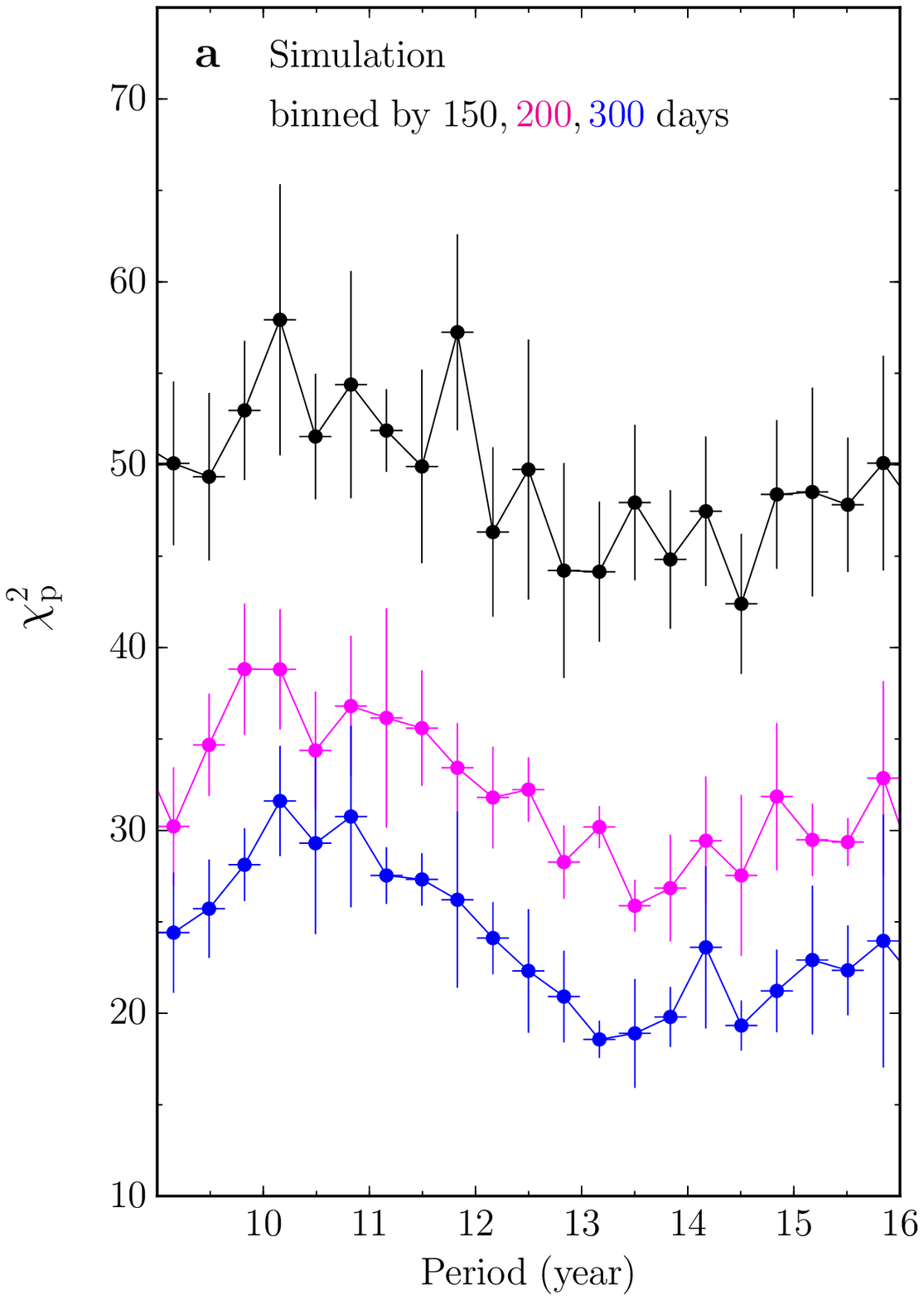}
\includegraphics[width=0.3\textwidth]{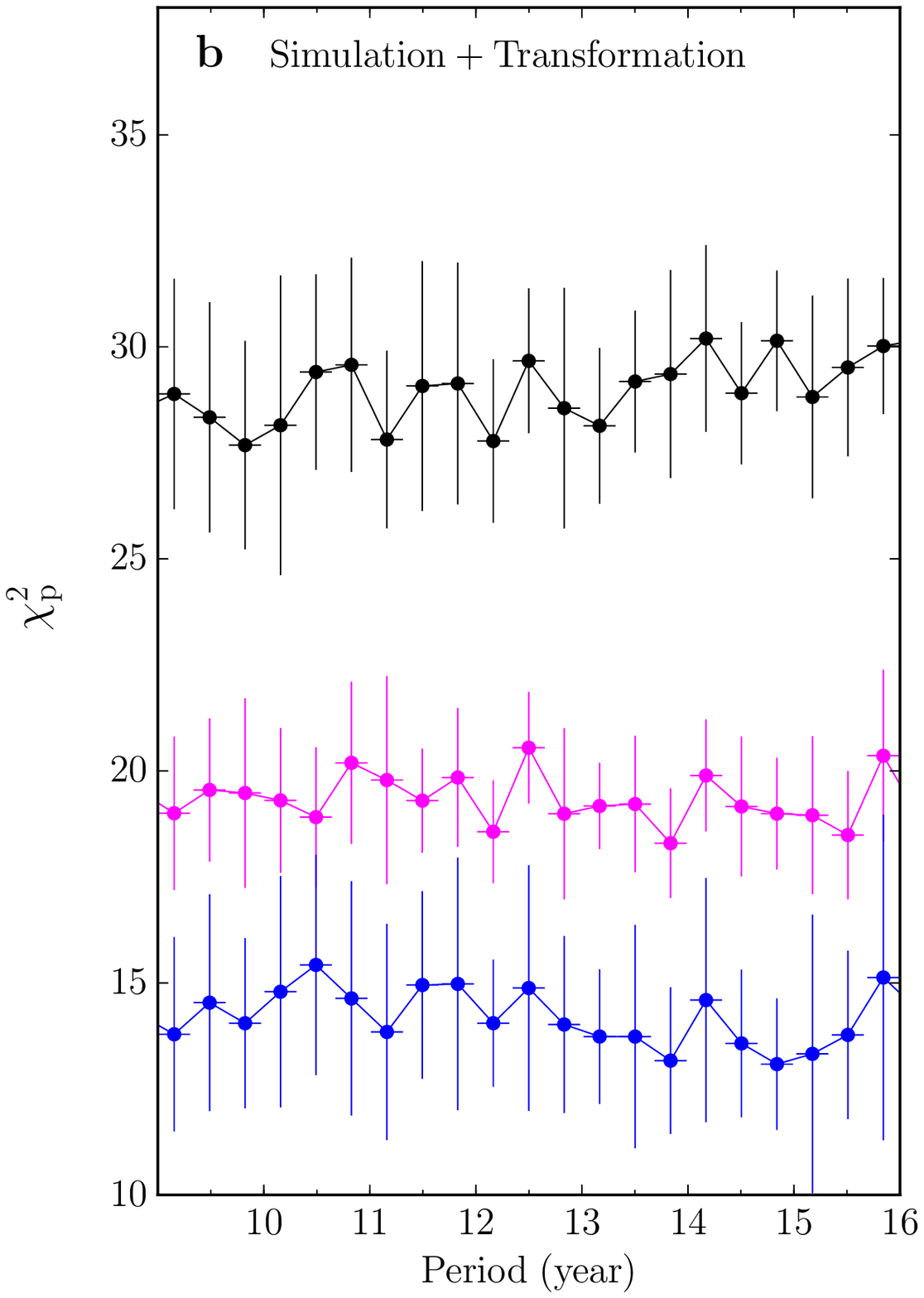}
\includegraphics[width=0.307\textwidth]{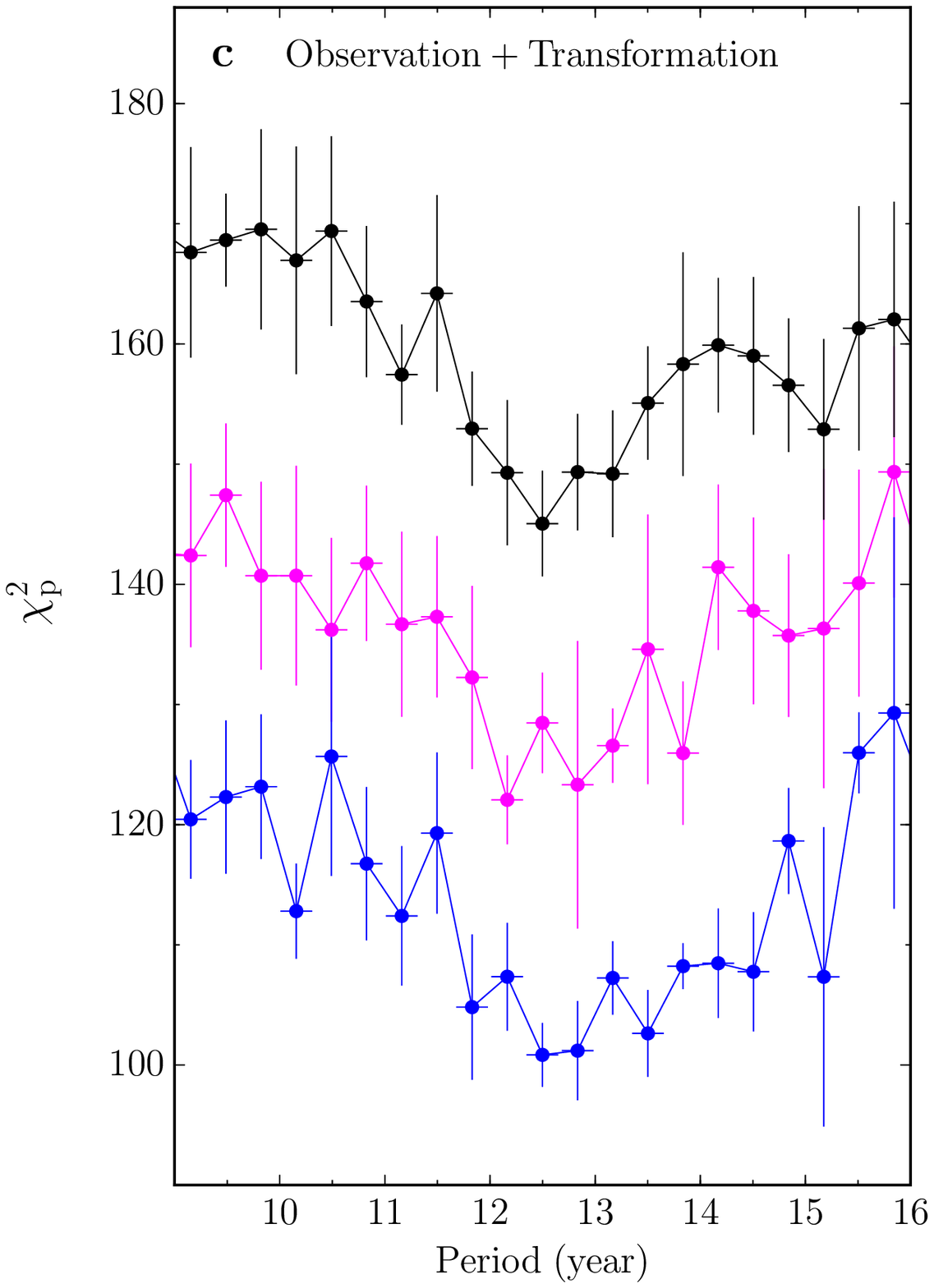}
\caption{Same as Figure~\ref{fig_phase} but for ({\it a}) simulation data, 
({\it b}) simulation data with a transformation, and ({\it c}) observation data
with a transformation (see the text for details).}
\label{fig_sim}
\end{figure}

\section{D. SMBHB orbital motion}
Figure~\ref{fig_sch} illustrates a schematic for the geometry of the SMBHB system. 
The observer is in the $XOZ$-plane, and the orbital plane is defined to be the $XOY$-plane, 
viewed at an inclination angle $i$. Given the total mass of the binary system $\bhm$, 
the orbital period $T_\bullet$ (rest-frame) 
and the orbital semi-major axis $a_\bullet$ are related by
\begin{equation}
a_\bullet= \left(\frac{G\bhm T_\bullet^2}{4\pi^2}\right)^{1/3}.
\end{equation}
The separation between the binary black holes is 
\begin{equation}
A_\bullet = \frac{a_\bullet(1-\bhe^2)}{1+\bhe \cos\theta},
\end{equation}
where $\bhe$ is the orbital eccentricity and $\theta$ is the angle from periastron, depending on the time as
\begin{equation}
\frac{d\theta}{dt}=\frac{2\pi}{T_\bullet}\frac{\left[1+\bhe\cos\theta(t+t_0)\right]^2}{(1+\bhe^2)^{3/2}},
\end{equation}
where $t_0=\phi_0T_\bullet$ is the time prior to the start of data-taking 
that the binary is at periastron.  By introducing the eccentric anomaly~$\xi$,
\begin{equation}
\sin\theta = \frac{\cos\xi -\bhe}{1-\bhe\cos \xi},
\end{equation}
we obtain a simple equation for $\xi$ (\citealp[p.38]{Hilditch2001})
\begin{equation}
\xi - \bhe\sin\xi = \frac{2\pi}{\bhT}(t + t_0).
\end{equation}
For an orbit with the periastron at an 
angle $\omega$ from the $X$-axis,  the positions of the primary and the 
secondary black holes are 
\begin{equation}
(x_1, y_1) = -\frac{q\BHA}{1+q}\left[\cos(\theta + \omega), \sin(\theta + \omega)\right],
\end{equation}
and 
\begin{equation}
(x_2, y_2) = \frac{\BHA}{1+q}\left[\cos(\theta + \omega), \sin(\theta + \omega)\right],
\end{equation}
respectively, where $q$ is the mass ratio of the binary.  Their corresponding velocities are 
\begin{equation}
(v_{x,1}, v_{y, 1}) = -\frac{q}{1+q}\left[v_r\cos(\theta + \omega)-v_\theta\sin(\theta + \omega),
v_r\sin(\theta + \omega)+v_\theta\cos(\theta + \omega)\right],
\end{equation}
and 
\begin{equation}
(v_{x,2}, v_{y, 2}) =\frac{1}{1+q}\left[v_r\cos(\theta + \omega)-v_\theta\sin(\theta + \omega),
v_r\sin(\theta + \omega)+v_\theta\cos(\theta + \omega)\right],
\end{equation}
where $v_r$ and $v_\theta$ are given by
\begin{equation}
v_r = \frac{dA_\bullet}{dt},~~~~~~v_\theta=A_\bullet\frac{d\theta}{dt}.
\end{equation}
The line-of-sight velocities of the binary are $V_{\rm LOS,1}=v_{x,1}\sin i$ 
and $V_{\rm LOS,2}=v_{x,2}\sin i$, respectively.

\end{document}